\documentclass[12pt,titlepage]{article}
\pdfoutput=1
\usepackage{graphicx}
\usepackage[letterpaper,hmargin=1in,vmargin=1.25in]{geometry}
\usepackage{times} 

\usepackage{comment}
\usepackage[usenames,dvipsnames]{color}
\newcommand{\commentcolor}[2]{\ifnum\Comments=1\textcolor{#1}{#2}\fi}
\usepackage{setspace} 
\usepackage{amsmath}
\usepackage{bm}
\usepackage{multirow}

\usepackage[round,numbers,sort&compress]{natbib}

\newcommand{\argmax}{\operatornamewithlimits{argmax}}
\newcommand{\argmin}{\operatornamewithlimits{argmin}}

\newcommand{\oh}{\frac{1}{2}}

\newcommand{\chem}[1]{\ensuremath{\mathrm{#1}}}
\renewcommand{\sup}[1]{\ensuremath{^{\mathrm{#1}}}}
\newcommand{\sub}[1]{\ensuremath{_{\mathrm{#1}}}}
\newcommand{\kclose}{\ensuremath{k_{\mathrm{close}}}}
\newcommand{\kopen}{\ensuremath{k_{\mathrm{open}}}}
\newcommand{\fsr}{FRET ratio}

\newcommand{\bz}{\{z\}_t}
\renewcommand{\bz}{{\bf z}}
\newcommand{\zhat}{{\hat\bz}}
\newcommand{\bzhat}{{{\bf{\hat\bz}}}}

\newcommand{\vbf}{ME}
\newcommand{\mlm}{ML}

\newcommand{\pztGyk}{p(\bz,\bt|\by,K)}

\newcommand{\pyzGtk}{p(\by,\bz|\bt,K)}
\newcommand{\ptGyk}{p(\bt|\by,K)}
\newcommand{\ptGk}{p(\bt|K)}
\newcommand{\ptsGk}{p(\ts|K)}
\newcommand{\pyGk}{p(\by|K)}
\newcommand{\pyGtk}{p(\by|\bt,K)}
\newcommand{\pyGt}{p(\by|\bt)}
\newcommand{\pyGtK}{p(\by|\bt,K)}
\newcommand{\pyztGk}{p(\by,\bt,\bz|K)}

\newcommand{\pzGtk}{p(\bz|{\bt},K)}

\newcommand{\qzt}{q(\bz,\bt)}
\newcommand{\qt}{q(\bt)}
\newcommand{\qzts}{q_*(\bz,\bt)}

\newcommand{\<}{\langle}
\renewcommand{\>}{\rangle}

\newcommand{\beq}{\begin{eqnarray}}
\newcommand{\eeq}{\end{eqnarray}}
\newcommand{\re}[1]{{\rm e}^{#1}}

\newcommand{\z}{\ensuremath{\bm z}}
\newcommand{\bx}{\ensuremath{\bm x}}
\newcommand{\bff}{\ensuremath{\bm f}}

\newcommand{\y}{\ensuremath{\bm y}}

\newcommand{\thv}{\ensuremath{\vec{\theta}}}

\newcommand{\kth}{\ensuremath{k^{th}}}

\newcommand{\btk}{\bt^K}
\newcommand{\bt}{{\vec{\theta}}}
\newcommand{\by}{{\bf{y}}}
\newcommand{\goto}{\rightarrow}
\newcommand{\infinity}{\infty}
\newcommand{\ts}{\bt_*}
\newcommand{\Ks}{K_*}
\newcommand{\dbtk}{d^K\bt}
\renewcommand{\theta}{\vartheta}

\newcommand{\upik}{\ensuremath{u_{\pi}^k}}
\newcommand{\uajk}{\ensuremath{u_{a}^{jk}}}
\newcommand{\umk}{\ensuremath{u_{\mu}^{k}}}
\newcommand{\ubk}{\ensuremath{u_{\beta}^{k}}}
\newcommand{\uvk}{\ensuremath{u_{v}^{k}}}
\newcommand{\uwk}{\ensuremath{u_{W}^{k}}}
\newcommand{\bupi}{\ensuremath{{\vec{u}_{\pi}}}}
\newcommand{\bua}{\ensuremath{\vec{u}_{a}}}
\newcommand{\bum}{\ensuremath{\vec{u}_{\mu}}}
\newcommand{\bub}{\ensuremath{\vec{u}_{\beta}}}
\newcommand{\buv}{\ensuremath{\vec{u}_{v}}}
\newcommand{\buW}{\ensuremath{\vec{u}_{W}}}

\newcommand{\cf}{{\it{cf.},} }
\newcommand{\ie}{{\it{i.e.},} }
\newcommand{\eg}{{\it{e.g.},} }

\newcommand{\fret}{smFRET}
\newcommand{\FRET}{{\small{\sc {FRET}}}}
\newcommand{\vbFRET}{{\small{\sc {vbFRET}}}}

\newcommand{\Eqn}{Eq. }

\newcommand{\flo}{f_{\rm low}}
\newcommand{\fhi}{f_{\rm high}}
\newcommand{\fmid}{f_{\rm mid}}
\newcommand{\fmidstate}{$\fmid$ state}

\title{Learning Rates and States from Biophysical Time Series: A 
Bayesian Approach to Model Selection 
and 
Single-Molecule FRET Data}

\author{Jonathan~E.~Bronson,$^* $
        Jingyi~Fei,$^*$
        Jake~M.~Hofman,$^\dagger$ \\
        Ruben~L.~Gonzalez~Jr.,$^*$
        ~Chris~H.~Wiggins$^\ddagger$ \thanks{
	    $^*$Department of Chemistry, Columbia University, 
	    3000 Broadway, New York, NY, 10027;
	    $^\dagger$Department of Physics, Columbia University,
	    538 W. 120th St., New York, NY, 10027. Present address:
	    Yahoo! Research, 111 West 40th Street
		New York, NY 10018;
        $^\ddagger$  
        Department of Applied Physics and Applied Mathematics,
        Columbia University, 200 S. W. Mudd Building,
	    New York, NY~10027.	  
}
}

\usepackage{layout}
\begin{document}
\maketitle

{\bf Abstract:} Time series data provided by single-molecule
F\"{o}rster resonance energy transfer (smFRET) experiments offer the
opportunity to infer not only model parameters 
describing molecular complexes,
e.g. rate constants, but also information about the model itself,
e.g. the number of conformational states. Resolving whether or how
many of such 
states exist requires a careful
approach to the problem of {\it model selection}, here meaning
discriminating among models 
with differing numbers of  states.
The most
straightforward 
approach to model selection generalizes the common
idea of maximum likelihood
---
selecting the {\it most likely parameter
values}
---
to maximum evidence:
selecting
the 
{\it most likely model}. 
In either case, 
such inference 
presents a tremendous computational challenge,
which we here 
address
by exploiting
an 
approximation 
technique 
termed 
{\it variational Bayes}.
We demonstrate how this
technique can be applied to temporal data such as
 smFRET time series; show superior statistical consistency
relative
to the maximum
likelihood approach; compare its performance
on smFRET data generated from experiments on the ribosome;
and illustrate how 
model selection in 
such probabilistic 
or generative 
modeling can facilitate analysis of closely related
temporal data currently prevalent in biophysics. 
Source code used in this analysis,
including a graphical user interface, is available open source via
{http://vbFRET.sourceforge.net.}

{\it Key words:}
{FRET;
smFRET;
Hidden Markov Model (HMM);
Model selection;
Variational Bayes;
statistical inference
}

\vfil\eject
\tableofcontents

\section{Introduction}
Single-molecule biology has triumphed at creating well-defined
experiments to analyze the workings of biological materials,
molecules, and enzymatic complexes. As the molecular machinery studied
become more complex, so too do the biological questions asked and,
necessarily, the statistical tools needed to answer these questions
from the resulting experimental data. In a number of recent
experiments, researchers have attempted to infer mechanical parameters
(e.g., the typical step size of a motor protein), probabilistic
parameters (e.g., the probability per turn that a topoisomerase
releases from its DNA substrate), or kinetic parameters (e.g., the
folding/unfolding rates of a ribozyme) via
statistical 
inference
\cite{koster2006mes, Bustamante09, Blanchard07,
  HerschlagChu00, ZhuangChu02, Gonzalez2008, Selvin04, Fernandez07,
  MyongHa05}.
Often the question of interest is not only one of selecting 
model parameters
but also 
selecting the model,
including from among models
which differ in the number of parameters to be inferred from
experimental data. The most
straightforward approach to model selection generalizes the common
idea of maximum likelihood (ML) --- selecting the {\it most likely
  parameter values} --- to maximum evidence (ME):
selecting
the 
{\it most likely model}. 

In this manuscript we focus on model selection in a specific example
of such a biological challenge: revealing the 
number of 
enzymatic conformational states
in single molecule FRET (smFRET) data.
FRET \citep{NatBioTechRev2003,JooHaRev2008, RoyHaRev2008,EatonRev2008}
refers to the transfer of energy from a donor fluorophore (which has
been excited by short-wavelength light) to an acceptor fluorophore
(which then emits light of a longer wavelength) with an efficiency
which decreases as the distance between the fluorophores
increases.  The distance-dependence of the energy transfer efficiency implies that
the quantification of the light emitted at both wavelengths from a
fluorophore pair may be used as a proxy for the actual distance
(typically $\sim$1--10 nm) between these fluorophores. Often a scalar
summary statistic (e.g. the ``FRET ratio'' $I_A/(I_A+I_D)$ of the
acceptor intensity to the sum of the acceptor and donor intensities)
is analyzed as a function of time, yielding time series data which are
determined by the geometric relationship between the two fluorophores
in a non-trivial way. When the donor and acceptor are biochemically
attached to a single molecular complex,
one may reasonably
interpret such a time series as deriving from the underlying conformational
dynamics of the complex.

If the complex of interest transitions from one locally stable
conformation to another, the experiment is well modeled by a hidden
Markov model (HMM) \citep{Rabiner1989}, a probabilistic model in which
an observed time series (here, the \fsr) is conditionally dependent on
a hidden, or unobserved,
discrete state variable
(here, the molecular conformation).
HMMs have long been used in ion channel experiments in which the
observed dynamic variable is voltage, and the hidden variable
represents whether the channel is open or closed
\citep{Sachs1997,Sachs2000}. More recently, Talaga proposed adapting
such modeling for FRET data \citep{Talaga03}, and Ha
and coworkers developed HMM software designed for FRET analysis \citep{HaMMy2006}.
Such existing software for biophysical time
series analysis implement ML on individual
traces and require users either to guess the number of states present in the
data, or to overfit the data intentionally by asserting an excess
number of states. Resulting errors 
commonly are then
corrected  via heuristics particular to each software package. It would be
advantageous to avoid subjectivity (as well as extra effort) on the
part of the experimentalist necessary in introducing thresholds or
other parameterized penalties for complex models, as well as to derive
principled approaches likely to generalize to new experimental
contexts and data types. To that end, our aim here is to implement
ME directly, avoiding overfitting even within the
analysis of each individual trace rather than as a post-processing
correction.

This manuscript begins by describing the general problem of using
probabilistic or {\it generative} models for experimental data
(generically denoted $\by$) in which one specifies the probability of
the data given a set of parameters of biophysical interest (denoted
$\bt$) and possibly some hidden value of the state variable of
interest (denoted $\bz$). We then present one particular framework,
variational Bayes, for estimating these parameters while at the same
time finding the optimal number of values for the hidden state
variable $\bz$. (In this manuscript bold variables are reserved for
those extensive in the number of observations.) We next validate the
approach on synthetic data generated by an HMM, with parameters
chosen to simulate data comparable to experimental smFRET data of
interest. Having validated the technique, we apply it to experimental smFRET
data and interpret our results. We close by highlighting advantages of
the approach; suggesting related biophysical time series data which
might be amenable to such analysis; and outlining promising avenues
for future extension and developments of our analysis.

\section{Parameter and model selection}

Since the techniques we present here are natural generalizations of
those which form the common introduction to statistical techniques in
a broad variety of natural sciences, we 
first
remind the reader of 
a few key ideas in inference
necessary 
before narrowing to the description of smFRET data, 
briefly discussing
ML methods for parameter inference and ME methods for
model selection. Extensions to temporal biophysical data more
generally will be discussed in Sec.~\ref{sec:conclusion}.  Note that, since the
ML-ME discussion does not rely on whether or not the model features
hidden variables, for the sake of simplicity we first describe in the
context of models without hidden variables.

  \subsection{Maximum likelihood inference}
  \label{sec:mlParams}
The context in which most natural scientists encounter statistical
inference is 
that of ML;
in this problem setting, 
the model is specified by
an expression for the {\it
  likelihood} $\pyGt$ --- i.e., the probability of the vector of data
$\by$ given some unknown vector of parameters of interest
$\bt$. (While this is not often stated explicitly, this is the
framework underlying minimization of $\chi^2$ or sums-of-squared
errors; {\it cf.}  Sec. \ref{sec:chisquared} for a less cursory
discussion.) In this context the ML estimate of the parameter $\bt$
is \beq
\label{eq:ML}
	\ts = \argmax_\bt \pyGt.
\eeq
ML  methods are useful for inference of parameter
settings under a fixed model (or model complexity), e.g. a particular
parameterized form with a fixed number of parameters. However, when one
would like to compare competing models (in addition to estimating
parameter settings), ML methods are generally
inappropriate, as they tend to ``overfit'', since likelihood increases
monotonically with model complexity.

This problem is conceptually illustrated in the case of inference from
FRET data as follows: if a particular system has a {\it known} number
of conformational states, say $K=2$, one can estimate the parameters
(the transition rates between states and relative occupation of states
per unit time) by maximizing the likelihood, which gives a formal
measure of the ``goodness of fit'' of the model to the data.
Consider, however, an overly complex model for the same observed data
with $K=3$ conformational states, which one might do if the number of
states is itself unknown. The resulting parameter estimates will have
a higher likelihood or ``better'' fit to the data under the maximum
likelihood criterion, as the additional parameters have provided more
degrees of freedom with which to fit the data. The difficulty here is
that maximizing the likelihood fails to accurately quantify the
desired notion of a ``good fit'' which should agree with past
observations, generalize to future ones and model the underlying
dynamics of the system. Indeed, consider the pathological limit in
which the number of states K is set equal to the number of FRET time
points observed.
The model will exactly match the observed FRET trace, but will
generalize poorly to future observations. It will have failed to model
the data at all and nothing will have been learned about the true
nature of the system; the parameter settings will simply be a
restatement of observations.

The difficulty in the above example is that one is permitted both to
select the model complexity (the number of parameters in the above
example) and to estimate single ``best'' parameter settings, which
results in overfitting. While there are several suggested solutions to
this problem (reviewed in \citep{Bishop2006,MacKay:2003}), we present
here a Bayesian solution for modeling FRET data which is both
theoretically principled and practically effective
(Sec.~\ref{sec:meModels}). In this approach, one extends the concepts
behind maximum likelihood to that of maximum {\it marginal}
likelihood, or {\it evidence}, which results in an alternative
quantitative measure of ``goodness of fit'' that explicitly penalizes
overfitting and enables one to perform model selection. The key
conceptual insight behind this approach is that one is prohibited from
selecting single ``best'' parameter settings for models considered,
and rather maintains probability distributions over {\it all}
parameter settings.

  \subsection{Maximum evidence inference}
  \label{sec:meModels}

The ML framework generalizes readily to the problem of
choosing among different models. This includes not only models of
different algebraic forms, but also among {\it nested} models in which
one model is a parametric limit of another, e.g. models with hidden
variables or in polynomial regression. (A two  state model is a
special case of a three  state model with an empty state;
a 
second order polynomial is a special case of a third order polynomial
with one coefficient set to 0.) In this case we introduce an index $K$
over possible models, e.g., the order of the polynomial to be fit or,
here, the number of conformational states, and hope to find the value
of $K_*$ which maximizes the probability of the data given the model,
$\pyGk$:
\beq \Ks =
\argmax_K \pyGk=\argmax_K \int d\vec{\theta} \pyGtK \ptGk.
        \label{eq:ME}
        \label{eqn:Kstardef}
\eeq 
The  quantity $\pyGk$ is referred to as the {\it marginal likelihood}, or {\it
evidence}, as unknown parameters are marginalized (or summed out) over
all possible settings. The second expression in Eq.~\ref{eq:ME} follows readily from the
rules of probability provided we are willing to model the parameters
themselves (in addition to the data) as random variables.  That is, we
must be willing to prescribe a distribution $\ptGk$ from which the
parameters are drawn given one choice of the model.  Since this term
is independent of the data $\by$, 
it is sometimes referred to as the {\it prior}; the treatment of
parameters as random variables is the one of the distinguishing
features of Bayesian statistics. (In fact, maximizing the evidence is
the principle behind the oft-used Bayesian information criterion
(BIC), an asymptotic approximation valid under a restricted set of
circumstances; {\it cf.}  Sec. \ref{sec:bic} for an intuition-building
derivation illustrating how ME prevents overfitting.)  In this form we
may interpret the marginal likelihood $\pyGk$ as an averaged version
of the likelihood $\pyGt$ over all possible parameter values, where
the prior $\ptGk$ weights each such value. Unlike the likelihood, the
evidence is largest for the model of correct complexity and decreases
for models that are either too simple or too complex with out the need
for any additional penalty terms. There are several explanations for
why evidence can be used for model selection
\citep{Bishop2006}. Perhaps the most intuitive is to think of the
evidence as the probability that the observed data was generated using
the given model (which we are allowed to do, since ME is a form of
generative modeling). Overly simplistic models cannot generate the
observed data and, therefore, have low evidence scores (e.g. it is
improbable that a two FRET state model would generate data with three
distinct FRET states). Overly complex models can describe the observed
data, however, they can generate so many different data sets that the
specific observed data set becomes improbable (e.g. it is improbable
that a 100 FRET state model would generate data that only has 3
distinct FRET states (especially when one considers that the evidence
is an average taken over all possible parameter values)).

In addition to performing model selection, we would like to make
inferences about model parameters, described by the
probability distribution over parameter settings given the
observed data, 
$\ptGyk$,
termed the {\it posterior}
distribution. 
Bayes' rule 
equates the posterior with
the 
product of the likelihood and the prior, normalized
by the evidence:
\beq 
\ptGyk= 
{\pyGtK \ptGk \over \pyGk}.
  \label{eq:pty}
  \eeq While ME  above does not give us access to the
posterior directly, as we show below, variational Bayes gives not only an
approximation to the evidence but also an approximation to the
posterior.

  \subsection{Variational approximate inference}
  \label{sec:variational_methods}
While in principle calculation 
of the
evidence and posterior completely
specifies the ME approach to model selection,
in practice exact computation of the evidence is
often both analytically and numerically intractable. 
One broad and intractable class is that arising
from models in which observed data are modeled
as conditionally dependent on an unknown or hidden
state to be inferred;
these {\it hidden
  variables} must be marginalized over (summed over) in calculating the
evidence in \Eqn~\ref{eq:ME}.
(For the smFRET data considered here,
these hidden variables represent the unobservable conformational
states.) As a result, calculation of
the evidence now involves a discrete sum over all states $\z$ in
addition to the integrals over parameter values \thv: \beq \pyGk=
\sum_{\z} \int d\theta \pyzGtk\ptGk.
\label{eq:MEz}
\eeq This significantly complicates the tasks of model selection and
posterior inference. 
Computing the terms in \Eqn \ref{eq:ME} and
\Eqn
\ref{eq:pty} requires calculation of the evidence, direct evaluation
of which requires a sum over all $K$ settings for each of $T$
extensive variables $\z$ (where $T$ is the length of the time series).
Such a sum is intractable for even $K=2$ and modest values of $T$,
e.g. on the order of 25. While there exist various methods for
numerically
approximating such sums, such as Monte Carlo techniques, we appeal
here to variational methods for a scalable, robust, and empirically
accurate method for approximate Bayesian inference. (For a discussion
regarding practical aspects of implementing Monte Carlo techniques,
including  burn-in, convergence rates, and
scaling, {\it cf.}  \cite{neal1993piu}.)

To motivate the variational method, we note that we wish not
only to select the model by determining $K_*$ but also to find
the posterior probability distribution for the parameters given the
data, i.e., $\pztGyk$. Variational Bayes amounts to finding the
distribution $\qzt$ which best approximates $\pztGyk$, i.e., 
\beq
\qzts=\argmin_{\qzt}D_{KL}\left({\qzt||\pztGyk}\right),  
\eeq
where $D_{KL}$ is the usual Kullback-Leibler divergence, which
quantifies the dissimilarity between two probability distributions.
A simple identity 
relates this quantity to the evidence $\pyGk$:
\beq
D_{KL}\left(\qzt||\pztGyk\right)
=\log \pyGk
+F[\qzt](\by)
\label{eq:DKL}
\eeq where the free energy $F[\qzt](\by)$ to be minimized, a function
of the data $\by$ and a functional of the test distribution $\qzt$, is
derived in Sec.~\ref{sec:dklproof}. Qualitatively, \Eqn~\ref{eq:DKL}
states that the log-evidence may be expressed as the difference
between an analytically tractable functional $F[\qzt]$ (owing to a
simple choice of the approximating distribution) and the dissimilarity
between the approximating distribution and the parameter posterior
distribution. Stated more succinctly:
the best test distribution $q$ not only gives the best estimate of the
evidence but also the best estimate of the posterior distribution of
the parameters themselves.
In going from \Eqn \ref{eq:MEz} to \Eqn \ref{eq:DKL},
we have replaced the problem of an intractable summation with that
of bound optimization. 

Calculation of $F$ is made tractable 
by choosing an approximating distribution $q$ with
conditional independence among variables which are 
coupled in the model given by $p$; for this reason the resulting technique 
generalizes mean field theory of statistical mechanics \cite{MacKay:2003}.
Just as in mean field theory,
the variational
method is defined by iterative update equations; here the update
equations result from setting the derivative of $F$ with respect to
each of the factors in the approximating distribution $q$ to $0$.  Since
$F$ is convex in each of these factors, the algorithm provably
converges to a local (though not necessarily global) optimum, and
multiple restarts are typically employed. Note that this is true for
expectation-maximization procedures more generally, including as
employed to maximize likelihood in models with hidden variables (e.g.,
HMMs). In ML inference, practitioners on occasion use the
converged result based on one judiciously chosen initial  condition
rather than choosing the optimum over restarts; this heuristic often
prevents pathological solutions ({\it cf.}
\citep{Bishop2006}, Ch. 9).

\section{Statistical inference and FRET}
  \subsection{Hidden Markov modeling}
  
The HMM \citep{Rabiner1989}, illustrated in
Fig.~\ref{fig:graphicalmodel},
models the dynamics of an observed time series $\by$ (here, the
observed
FRET ratio) 
as conditionally dependent 
on a hidden process $\bz$ (here, the unknown conformational state
of the molecular complex). At each time $t$, the conformational state
$z_t$
can take on any one of $K$ possible values, conditionally dependent
only on its value at the previous time via the transition probability
matrix $p(z_t|z_{t-1})$ (i.e., $\bz$ is a Markov process); 
the observed data depend only on the current-time
hidden state via the emission probability $p(y_t|z_t)$.
Following
the convention to the field, we model the emission probability $p(y_t|z_t)$ as
a Gaussian \citep{HaMMy2006,Weiss1999}, ignoring for the moment the complication
of modeling a variable distributed on the interval $[0,1]$ with a distribution
of support $(-\infty,\infty)$.

For \fret\ time series with observed 
data
$(y_1,\ldots,y_T) =
\y$ and corresponding hidden state conformations
$(z_1,\ldots,z_T) = \z$, the joint probability of the observed and
hidden data
is 
\beq p(\y,\z|\bt,K) = p(z_{1}|\bt)
\left[ \prod_{t=2}^{T}
p(z_{t}|z_{t-1},\bt)\right]\prod_{t=1}^{T}
p(y_t|z_{t},\bt) 
\label{eq:hmm}
\eeq where $\bt$ comprises four types of parameters: a $K$-element
vector, $\vec{\pi}$ where the $k^{th}$ component, $\pi_k$, holds the
probability of starting in the \kth\ state; a $K\times K$ transition
matrix, $A$, where $a_{ij}$ is the probability of transitioning from
the $i^{th}$ hidden state to the $j^{th}$ hidden state (i.e. $a_{ij} =
p(z_{t}=j|z_{t-1}=i)$); and two $K$-element vectors, $\vec{\mu}$ and
$\vec{\lambda}$, where $\mu_k$ and $\lambda_k$ are the mean and precision
of the Gaussian distribution of the \kth\ state.

As in Eq.~\ref{eq:MEz}, the evidence follows directly from multiplying the likelihood
by priors and marginalizing:
 
\beq p(\y|K) = \sum_{\z} \int d\bt
p(\vec{\pi})p(A)p(\vec{\mu},\vec{\lambda}) p(z_{1}|\vec{\pi}) \left[
  \prod_{t=2}^{T} p(z_{t}|z_{t-1},A)\right] \prod_{t=1}^{T}
p(y_t|z_{t},\vec{\mu},\vec{\lambda}).
\label{eq:hmm_evidence}
\eeq The $p(\vec{\pi})$ and each row of A are modeled as Dirichlet
distributions; each pair of $\mu_k$ and
$\lambda_k$ are modeled jointly as a Gaussian-Gamma
distribution. 
Algebraic 
expressions for these distributions can be
found in the Supporting Material (Sec.~\ref{sec:prior_math}). Their
parameter settings and the effect of their parameter settings on data
inference can be found in Sec.~\ref{sec:hyperparameters} and
Sec.~\ref{sec:hpar_results}, respectively. We found that for the
experiments considered here, and the range of prior parameters  tested,
there is little discernible effect of the prior parameter settings on
the data  inference.   
The variational 
update equations
approximating
Eq.~\ref{eq:hmm_evidence} with these priors can be found in
\citep{Carin2006}.

The variational approximation to the above evidence 
utilizes the dynamic program termed the
forward-backward algorithm \citep{Rabiner1989}, 
which 
requires O($K^2T$) computations, 
rendering the computation feasible. (In comparison, 
direct summation
over all terms requires $O(K^T)$ operations.) We
emphasize that, while individual steps in the ME calculation are
slightly more expensive than their ML counterparts, the scaling
with the number of states and observations is identical.  As discussed in
section \ref{sec:variational_methods}, in addition to calculating the
evidence the variational solution yields a 
distribution
approximating
the probability of the
parameters given the data. Idealized traces can be calculated by
taking the most probable parameters from these distributions and
calculating the most probable hidden state trajectory using the
Viterbi algorithm \citep{viterbi}.

  \subsection{Rates from states}
  HMMs are used to infer the number of conformational states present in
the molecular complex as well as the transition rates between
states. Here, we follow the convention of the field by fitting every
trace individually (since the number and mean values of \fret\ states
often 
vary 
from traces to trace). Unavoidably then, an ambiguity is
introduced 
comparing
FRET state labels
across multiple traces, since ``state 2'' may refer to
the high variant of a low state in one trace and to the low variant of
a high state in a separate trace.  To overcome this 
ambiguity, rates are not inferred directly
from $\qt$, but rather from the idealized traces $\zhat$ where 
\beq
\zhat=\argmax_{\bz}{q(\bz|\by,\thv_\dagger,K)} 
\label{eqn:zhatdef}
\eeq 
and $\thv_\dagger$
are, for ME, the 
parameters specifying the optimal parameter distribution $q_*(\thv,\bz)$
or, for ML, the most likely parameters, \thv$_*$.  The number of
states in the data set can then be determined by combining the
idealized traces and plotting a 1D FRET histogram or transition
density plot (TDP).  
Inference facilitates the calculation of
transition rates 
by, for example,
dwell-time analysis, TDP analysis, or by dividing the sum of the dwell
times by the total number of transitions
\citep{NollerHa2008,HaMMy2006}. In this work, we determine the number
of states in an individual trace using ME. To overcome the ambiguity
of labels when combining traces, we follow the convention of the field
and use 1D FRET histograms and/or TDPs to infer the number of states
in experimental data sets and calculate rates using dwell time
analysis (Sec.~\ref{sec:rates_methods}).

\section{Numerical experiments}

  We created a software package to implement variational Bayes for FRET
data called \vbFRET. Software was written in {\sc{MATLAB}} and is available
open source, including a point and click GUI. All ME data inference was
performed using \vbFRET. All ML data inference was performed using HaMMy
\citep{HaMMy2006}, although we note that any analysis
based on ML should perform similarly (see Sec.~\ref{sec:ml-restarts}
for practicalities regarding implementing ML). Parameter settings used for both
programs, methods for creating computer generated synthetic data, and
methods for calculating rate constants for experimental data can be found in
Sec.~\ref{sec:methods}.
Following the convention of the field, in subsequent sections
the dimensionless FRET ratio is quoted in dimensionless ``units" of \FRET.

  \subsection{Example: maximum likelihood vs maximum evidence}
  To illustrate the differences between ML and ME, consider the
synthetic trace shown in Fig.~\ref{fig:overfit}, generated with three
noisy states ($K_0=3$) centered at $\mu_z$ = (0.41, 0.61, 0.81) \FRET.
This trace was 
analyzed 
by both 
\vbf\ and \mlm\
with 
$K = 1$ (underfit), $K = 3$ (correctly fit), and $K = 5$ (overfit)
(Fig.~\ref{fig:overfit}{\it A}). In the cases when only
one or three states are allowed, 
\vbf\ and \mlm\ perform similarly.
However, when five states are allowed, \mlm\ overfits the data, whereas
\vbf\ leaves two states unpopulated and correctly infers three states,
illustrated clearly via the idealized trace.

Moreover, whereas the likelihood of the overfitting model
is larger than that of the correct model,
the evidence is largest when
only three states are allowed 
($p(\by|\ts,K>K_0)>p(\by|\ts,K_0)$; however, $p(\by|K)$ peaks
at $K=K_0=3$).
The ability to use the evidence for model selection is further
illustrated in Fig.~\ref{fig:overfit}{\it B}, in which
the same data as in 
Fig.~\ref{fig:overfit}{\it A} 
are
analyzed using both
\vbf\ and \mlm\ with $1 \leq K \leq
10$. 
The evidence is greatest when $K = 3$; however, the likelihood
increases monotonically as more states are allowed, ultimately
leveling off after five or six states are allowed.

  \subsection{Statistical validation}
  ME can be statistically validated
by generating synthetic data, for which
the true trajectory of the hidden state $\bz_0$ is known,
and quantifying performance relative to ML.
We performed such numerical experiments,
generating several thousand synthetic traces,
and quantified accuracy as a function of signal-to-noise via
four probabilities:
(1) accuracy in number of states $p(|\bzhat|=|\bz_0|)$: 
the probability in any trace of inferring the correct number of states
(where 
$|\bz_0|$ is the number of 
states in the model generating the data and
$|\bzhat|$ is the number of populated states in the idealized trace);
(2) accuracy in states $p(\bzhat=\bz_0)$:
the probability in any trace at any time of inferring the correct state;
(3) sensitivity to true transitions: 
the probability in any trace at any time that the
inferred trace $\bzhat$ exhibits a transition, 
given that $\bz_0$ does; and
(4) specificity of inferred transitions: 
the probability in any trace at any time that the
true trace $\bz_0$ does not exhibit a transition, 
given that $\bzhat$ does not.
We note that, 
encouragingly, for the ME inference, $|\bzhat|$ always equaled $K_*$ as defined in \Eqn \ref{eqn:Kstardef}.

We identify each inferred
state with the true state which is closest in terms of their means 
provided the difference in means is less than $0.1$ \FRET. Inferred
states for which no true state is within $0.1$ \FRET\ are considered
inaccurate. Note that we do not demand that one and only one inferred
state be identified with the true state. This effective smoothing
corrects
overfitting errors in which one true state has been inaccurately
described by two nearby states (consistent with the convention of the
field for analyzing experimental data).

For all synthetic traces, $K_0$ = 3 with means centered at $\mu_z$ =
(0.25, 0.5, 0.75) \FRET. Traces were made increasingly noisy by
increasing the standard deviation, $\sigma$, of each 
state.
Ten different noise levels, ranging from $\sigma \approx 0.02$
(unrealistically noiseless) to $\sigma\approx 1.4$ 
(unrealistically noisy) were
used. Trace length, $T$, varied from $50 \leq T \leq 500$ time steps,
drawn randomly from a uniform distribution. One time step corresponds to one
time-binned unit of an experimental trace, which is typically 25--100 msec for
most CCD camera based experiments.  Fast-transitioning (mean lifetime
of 4 time steps between transitions) and slow-transitioning (mean
lifetime of 15 time steps between transitions) traces were created and
analyzed separately. Transitions were equally likely from all hidden
states to all hidden states. For each of the 10 noise levels and 2 transition
speeds, 100 traces were generated (2,000 traces in total). Traces
for which $K_0$ = 2 (Fig.~\ref{fig:synthetic_results2}) and $K_0$ = 4
(Fig.~\ref{fig:synthetic_results4}) were created and analyzed as
well. The results were qualitatively similar and can be found in
Sec.~\ref{sec:validation_sup}.

As expected, both programs performed better on low noise traces than
on high noise traces. \vbf\ correctly determined the number of FRET
states more often than \mlm\ in all cases except for the noisiest
fast-transitioning trace set (Fig.~\ref{fig:gauss3results}, top left).
Of the 2,000 traces analyzed here using \vbf\ and \mlm,
\vbf\ overfit 1 and underfit
232. \mlm\ overfit 767 and underfit 391. 
In short, ME essentially eliminated 
overfitting of the individual traces, whereas ML
overfit 38\% of individual traces.
Over 95\% (all but 9) of \vbf\ underfitting errors occurred on traces
with FRET state noise $>$ 0.09, whereas \mlm\ underfitting was much more
evenly distributed (at least 30 traces at every noise level were
underfit by \mlm). The underfitting of noisy traces by \vbf\ may be a
result of the intrinsic resolvability of the data, rather than a
shortcoming of the inference algorithm; as the noise of two adjacent
states becomes much larger than the spacing between them, the two
states become indistinguishable from a single noisy state (in the
limit, there is no difference between a one state and two state system
if the states are infinitely noisy). The causes of the underfitting
errors by \mlm\ are less easily explained, but suggest that the \mlm\
algorithm  has not converged to a global optimum in
likelihood (for reasons explained in Sec.~\ref{sec:ME_settings}).

In analyzing
the slow-transitioning traces, the methods performed roughly
the same on probabilities (2--4) (always within $\sim$5\% of each
other). For the fast-transitioning traces, however, \vbf\ was much better
at inferring the true trajectory of traces (by a factor of 1.5--1.6
for all noise levels) and showed superior sensitivity (factor of
2.7--12.5) to transitions at all noise levels. The two methods showed
the same specificity to transitions until a noise level of $\sigma >$
0.8, beyond which \mlm\ showed better specificity (factor of
1.06--1.13). 
Inspection of the individual traces showed that
all three of these results were due to \mlm\ missing many of the
transitions in the data.

These 
results 
on synthetic data
suggest that when the number of states in the
system is unknown, \vbf\ clearly performs better at identifying FRET
states.
For inference of idealized trajectories, \vbf\ is at least as accurate as ML
for slow-transitioning traces and more accurate for
fast-transitioning traces. The performance of \vbf\ on fast-transitioning
traces is particularly
encouraging since detection of a transient biophysical state is often
an important objective of smFRET experiments, as discussed below.

\section{Results}
\label{sec:results}
Having validated inference with \vbFRET,
we 
compared ME and ML inference
on experimental smFRET data,
focusing our attention on the number of states and the 
transition rates.
The data we
used for this analysis report on the conformational dynamics of the
ribosome, the universally-conserved ribonucleoprotein enzyme
responsible for protein synthesis, or translation, in all
organisms. One of the most dynamic features of translation is the
precisely directed mRNA and tRNA movements that occur during the
translocation step of translation elongation. Structural, biochemical,
and smFRET data overwhelmingly support the view that, during this
process, ribosomal domain rearrangements are involved in directing
tRNA movements \citep{Noller89, BlanchardChu04, Blanchard07, Noller07,
  KimChu07, Gonzalez2008, NollerHa2008, Frank08, Valle08,
  NollerHa09}. One such ribosomal domain is the L1 stalk, which
undergoes conformational changes between open and closed conformations
that correlate with tRNA movements between so-called classical and
hybrid ribosome-bound configurations (Fig.~\ref{fig:ribosome}{\it A})
\citep{Gonzalez2008, NollerHa09, Fei09, Sam09}.

Using fluorescently-labeled tRNAs and ribosomes, we have recently
developed smFRET probes between tRNAs (smFRET\chem{_{tRNA-tRNA}})
\citep{BlanchardChu04}, ribosomal proteins L1 and L9
(smFRET\chem{_{L1-L9}}) \citep{Gonzalez2008}, and ribosomal protein L1
and tRNA (smFRET\chem{_{L1-tRNA}}) \citep{Fei09}. Collectively, these data
demonstrate that, upon peptide bond formation, tRNAs within
pretranslocation (PRE) ribosomal complexes undergo thermally-driven
fluctuations between classical and hybrid configurations
(smFRET\chem{_{tRNA-tRNA}}) that are
coupled to transitions of the L1 stalk between open and closed
conformations (smFRET\chem{_{L1-L9}}). The net result of these
dynamics is the transient formation of a direct L1 stalk-tRNA contact
that persists until the tRNA and the L1 stalk stochastically fluctuate
back to their classical and open conformations, respectively
(smFRET\chem{_{L1-tRNA}}). This intermolecular L1 stalk-tRNA contact
is stabilized by binding of elongation factor G (EF-G) to PRE and
maintained during EF-G catalyzed translocation \citep{Gonzalez2008,Fei09}.

Here we compare the rates of L1 stalk closing (\kclose) and opening
(\kopen) obtained from \vbf\ and \mlm\ analysis of smFRET\chem{_{L1-L9}} PRE
complex analogs (PMN) under various conditions (which have the same number of FRET
states by both inference methods) and the number of states inferred
for smFRET\chem{_{L1-tRNA}} PMN complexes by \vbf\ and \mlm. (FRET complexes
shown in Fig.~\ref{fig:ribosome}{\it B}.)  These data were chosen for
their diversity of sm\fsr s. The smFRET\chem{_{L1-L9}} ratio
fluctuates between FRET states centered at 0.34 and 0.56 (i.e. a
separation of 0.22 \FRET), whereas the smFRET\chem{_{L1-tRNA}} ratio
fluctuates between FRET states centered at 
$0.09$ and 
$0.59$ \FRET (i.e. a
separation of 0.50 \FRET). In addition, smFRET\chem{_{L1-L9}} data were
recorded under conditions that favor either fast-transitioning
(PMN\chem{_{fMet+EFG}}) or slow-transitioning (PMN\sub{fMet} and
PMN\sub{Phe}) complexes (complex compositions listed in
Table~\ref{table:real_dat}).

First, we compared the smFRET\sub{L1-L9} data obtained from
PMN\sub{fMet}, PMN\sub{Phe}, and PMN\sub{fMet+EFG}. As expected from
previous studies \citep{Fei09}, 1D histograms of idealized \FRET
values from both inference methods showed two \FRET states centered at
$ 0.34$ and $0.56$ \FRET\ 
(and one additional
state due to photobleaching, for a total of three states). When
individual traces were examined for overfitting, however, \mlm\ inferred
four or five states in $20.1\% \pm 3.7\%$ of traces in each data set
whereas \vbf\ inferred four or five states in only $0.9\% \pm 0.5$ of
traces. Consequently, more post-processing was necessary to extract
transition rates from idealized traces inferred by \mlm.

Our results (Table~\ref{table:real_dat}) demonstrate that there is
very good overall agreement between the values of \kclose\ and \kopen\
calculated by \vbf\ and \mlm. For the relatively slow-transitioning
PMN\sub{fMet} and PMN\sub{Phe} data, the values of \kclose\ and
\kopen\ obtained from \vbf\ and \mlm\ are indistinguishable. For the
relatively fast-transitioning PMN\sub{fMet+EFG} data, however, the
values of \kclose\ and \kopen\ obtained differ slightly between \vbf\ and
\mlm. Since the true transition rates of the experimental smFRET\sub{L1-L9} data
can never be known, it is impossible to assess the accuracy of the
rate constants obtained from \vbf\ or \mlm\ in the same way we could with
the analysis of synthetic data. While we cannot say which set
of \kclose\ and \kopen\   values are most accurate
for this fast-transitioning data set, our synthetic results would
predict a larger difference between rate constants calculated by \vbf\
and \mlm\ for faster-transitioning data and suggest that the values of
\kclose\ and \kopen\ calculated with \vbf\ have higher accuracy
(Fig.~\ref{fig:gauss3results}).

Consistent with previous reports \citep{Gonzalez2008}, \mlm\ infers two
FRET states centered at 
$\flo\equiv 0.09$
and 
$\fhi\equiv 0.59$ \FRET\  
(plus one photobleached state) for all smFRET\sub{L1-tRNA} data sets. 
Conflicting with these
results, however, \vbf\ infers three FRET states
(plus a photobleached state) for these data sets. Two of these FRET states are centered at
$\flo$ and $\fhi$, as in the \mlm\ case, while the third ``putative'' state
is centered at $\fmid\equiv 0.35$ \FRET, coincidentally at the mean between $\flo$ and $\fhi$. 
Indeed, TDPs constructed from the idealized trajectories generated by
\vbf\ or \mlm\ analysis of the PMN\sub{fMet+EFG} smFRET\sub{L1-tRNA} data
set evidence
the appearance of a new, highly populated state at $\fmid$ in the
\vbf-derived TDP that is virtually absent in the \mlm-derived TDP
(Fig.~\ref{fig:TDPs}). Consistent with the TDPs, $\sim$46\% of
transitions in the \vbf-analyzed smFRET\sub{L1-tRNA} trajectories are
either to or from the new $\fmid$ state (Fig.~\ref{fig:blur}{\it
  B}). This $\fmid$ state is extremely short lived; $\sim$75\% of the
data assigned to $\fmid$ consist of a single observation, \ie with a
duration at or below the CCD integration time (here, 50 msec)
(Fig.~\ref{fig:blur}{\it C}). A representative \vbf-analyzed
smFRET\sub{L1-tRNA} trace is shown in Fig.~\ref{fig:blur}{\it A}.

There are at least two possible origins for this putative new state.
The first 
is a 
very short-lived (i.e. lifetime $\leq$ 50 msec), {\it
  bona fide}, 
previously unidentified intermediate conformation of the PMN complex. The second 
is that 
$\fmid$
data are artifactual, resulting from
the 
binning of the continuous-time FRET signal during CCD collection.
Each
time binned data point represents the average intensity of thousands
or more photons. If a transition occurs 25 msec into a 50 msec time
step, half the photons will come from the 
$\flo$
state and half
from the 
$\fhi$ state,
resulting in a datum at approximately their mean.
This type of CCD blurring artifact would be
lost in the noise of closely spaced FRET states, but would become more
noticeable as the FRET separation between states increases.
 
To distinguish between these two possibilities, we recorded
PMN\sub{fMet+EFG} smFRET\sub{L1-tRNA} data at half and double the
integration times (i.e. 25 msec and 100 msec). If the 
$\fmid$ state
is a true conformational intermediate then: (1) the percentage of
transitions exhibiting at least one data point at or near 
$\fmid$
should increase as the integration time
decreases, and (2) the number of consecutive data points defining the
dwell time spent at or near 
$\fmid$
should increase as the
integration time decreases. Conversely, if the 
$\fmid$ state
arises from a time averaging artifact, then: (1) the percentage of
transitions containing at least one data point at or near 
$\fmid$
should increase as the integration time increases, as
longer integration times increase the probability that a transition
will occur during the integration time, and (2) the number of
consecutive data points defining the dwell time spent at or near the
\fmidstate\ 
should be independent of the integration time,
as transitions occurring within the integration time will always be
averaged to generate a single data point.

Consistent with the view that the 
\fmidstate\
arises from time-averaging over the integration time, Fig.~\ref{fig:blur}{\it B}
demonstrates that the percentage of transitions containing at least
one data point at or near 
$\fmid$
 increases as the
integration time increases. This manifests as the increase in the
density of transitions starting or ending at $\fmid$ as the
integration time decreases for the \vbf-derived TDPs in
Fig.~\ref{fig:TDPs}. These data are further supported by the results
presented in Fig.~\ref{fig:blur}{\it C}, demonstrating that the number
of consecutive data points defining the dwell time of the \fmidstate\ is
remarkably insensitive to the integration time. 
We conclude that the \fmidstate\ identified by \vbf\ is composed
primarily of a time-averaging artifact which we refer to as ``camera
blurring'' and the ME-inferred \fmidstate\ as the ``blur
state''. Although \mlm\ infers four or five states in 35\% of the
traces (compared to only 25\% for ME), for some reason \mlm\
significantly suppresses, but does not completely eliminate, detection
of this blur state in the individual smFRET trajectories. At present,
we cannot determine whether this is a result of the ML method itself
(i.e. overfitting noise in one part of the trace may cause it to miss
a state in another) or due to the specific implementation of ML in the
software we used (Sec.~\ref{sec:ML_setting}).  In retrospect, the
presence of blur states should not be surprising, since they follow
trivially from the time-averaging that results from averaging over the
CCD integration time.

Single data point artifacts caused by stochastic photophysical
fluctuations of fluorophore intensity are a well known and common
problem in smFRET data \citep{RoyHaRev2008}. These artifacts can be
corrected for by applying smoothing algorithms or rolling averages
over the data \citep{BlanchardChu04, NollerHa09} or
  ignoring FRET states with a dwell time of one time point
  \citep{Gonzalez2008}. The artifacts we encounter here are different in
nature, since they result from time binning the data rather than a
photophysical fluctuation in donor/acceptor signal intensity and,
therefore, should be corrected for using a different approach. The
algorithm we propose performs a second round of ME inference on the
data, using the idealized traces from the first round of \vbf\ inference
to make the following modification to the raw data: data which could
have resulted from time-averaging artifacts (i.e. events lasting
exactly one data point and occurring between two distinct idealized
values) were moved to the idealized value closest to the value of the
suspected time-averaging artifact (the assumption here is that that a
single $\fmid$ data point should be considered part of
the ``real'' FRET state that the molecular complex spent the most time
in during that transitioning time point).  We performed this algorithm
on the smFRET\sub{L1-tRNA}. The TDP for this ``cleaned'' data shows
the blur state at $\fmid$ is virtually eliminated, yielding a result
that is wholly consistent with that generated by \mlm\
(Fig.~\ref{fig:TDPs}). In general, however, it should be cautioned
that a {\it bona fide} intermediate FRET state may well exist and be
buried under a strongly-populated blur state. Unless this intermediate
FRET state is positively identified and somehow separated from the
blur state (i.e. by obtaining data at an increased integration time),
eliminating or ignoring FRET states with dwell times exactly equal to
one time point may risk overlooking a {\it bona fide} intermediate
FRET state. We note that the \vbFRET\ software package which we have
made available allows the user the opportunity to run this second
round of ME analysis with possible blur states detected and cleaned as
described above.

The observation that \mlm\ analysis does not detect a blur
state that is readily identified by \vbf\ analysis is in line with our
results on synthetic data in which \vbf\ consistently outperforms \mlm\ in
regards to detecting the true number of states in the data,
particularly in fast-transitioning data, and strongly suggests that \vbf\
will generally capture short-lived intermediate FRET states that \mlm\
will tend to overlook. While this feature of \mlm\ might be desirable in
terms of suppressing blur states such as the one we have identified in
the smFRET\sub{L1-tRNA} data set, it is undesirable in terms of
detecting {\it bona fide} intermediate FRET states that may exist in a
particular data set.

\section{Conclusions}
\label{sec:conclusion}
These synthetic and experimental analyses
confirm that ME can be used for model selection (identification
of the number of smFRET states) at the level of
individual traces, improving accuracy and avoiding overfitting.
Additionally, ME inference with variational Bayes
provides $q_*$, an estimate of the
true parameter and idealized trace posterior, making possible
the analysis of kinetic parameters, again at the level of individual
traces.
 As a tool for inferring idealized traces, \vbf\ produces traces
 which are visually similar to those of \mlm;
in the
 case of synthetic data generated to emulate experimental data,
\vbf\ performs with comparable or superior accuracy.
The idealized trajectories 
inferred 
by
\vbf\ required substantially less post-processing, however, since \vbf\
usually inferred the correct number of states to the data and,
consequently, did not require states with similar idealized values
within the same trace to be combined in a post-processing step. The
superior trajectory inference, accuracy, and sensitivity to transitions
of \vbf\ on fast transitioning synthetic traces suggests that the
differences in transition rates calculated for fast transitioning experimental
data is a result of superior fitting by \vbf\ as well.

In some experimental data, \vbf\ detected a very short lived blur
state, which comparison of experiments at different
sampling rates suggests results from a camera time averaging
artifact. Once detected by \vbf, the presence of this intermediate state
is easily confirmed by visual inspection, but yet was not identified by \mlm\
inference. Although not biologically relevant in this instance, this
result suggests that \vbf\ inference is able to uncover  real
biological intermediates in smFRET data that would be missed by \mlm.

We conclude by emphasizing that this method of data inference is in no
way specific to smFRET. The use of ME and variational Bayes could
improve inference for other forms of biological time series where the
number of molecular conformations is unknown. Some examples include
motor protein trajectories with an unknown number of chemomechanical
cycles (i.e. steps), DNA/enzyme binding studies with an unknown number
of binding sites and molecular dynamics simulations where important
residues exhibit an unknown number of rotamers.

All code used in this analysis, as well as a point and click GUI
interface, is available open source via http://vbFRET.sourceforge.net.

\section{Acknowledgments}
It is a pleasure to acknowledge helpful conversations with Taekjip Ha
and Vijay Pande, Harold Kim and Eric Greene for comments on the
manuscript, mathematical collaboration with 
Alexandro D. Ramirez,
and Subhasree Das for managing the Gonzalez laboratory. This work was
supported by a grant to CHW from the NIH (5PN2EY016586-03) and grants
to RLG from the Burroughs Wellcome Fund (CABS 1004856), the NSF (MCB
0644262), and the NIH-NIGMS (1RO1GM084288-01).

\clearpage
\begin{table}[h]
 
  \caption{Comparison of smFRET\sub{L1-L9} transition rates inferred by \vbf\ and \mlm}
  \begin{tabular}{l c c c }
    & & & \\
    Data set$^*$ & Method & \kclose & \kopen \\ \hline \hline

    &&& \\[-3.5mm]

    PMN\sub{Phe}$^{\dagger}$ & \vbf\ &  $0.66 \pm 0.05$ & $1.0 \pm 0.2$ \\

    & \mlm\ & $0.65 \pm 0.06$ & $1.0 \pm 0.3$\\ \hline 

    &&& \\[-3.5mm]

    PMN\sub{fMet}$^{\ddagger}$ & \vbf\ &  $0.53 \pm 0.08$ & $1.7 \pm 0.3$ \\

    & \mlm\  & $0.52 \pm 0.06$ & $1.8 \pm 0.3$ \\ \hline

    &&& \\[-3.5mm]

    PMN\sub{fMet+EFG}  & \vbf\ &  $3.1 \pm 0.6$ &
    $1.3 \pm 0.2$ \\

     ($1\mu M$)$^{\S}$ & \mlm\ & $2.1 \pm 0.4$ & $1.0 \pm 0.2$ \\\hline

    &&& \\[-3.5mm]

    PMN\sub{fMet+EFG}  & \vbf\ & $2.6 \pm 0.6$ & $1.5 \pm 0.1$ \\

     ($0.5\mu M$)$^{\S}$ & \mlm\ & $2.0 \pm 0.3$ & $1.0 \pm 0.1$ \\\hline

 \end{tabular}
 \label{table:real_dat}
\end{table}

\newcommand{\tempf}{\noindent \footnotesize}

{\tempf $^*$ Rates reported here are the average and standard deviation
  from three or four independent data sets. Rates were not corrected
  for photobleaching of the fluorophores.}

{\tempf $^\dagger$ PMN\sub{Phe} was prepared by adding the antibiotic
  puromycin to a post-translocation complex carrying
  deacylated-tRNA\sup{fMet} at the E site and fMet-Phe-tRNA\sup{Phe}
  at the P site, and thus contains a deacylated-tRNA\sup{Phe} at the P
  site.}

{\tempf $^\ddagger$ PMN\sub{fMet} was prepared by adding the antibiotic
  puromycin to an initiation complex carrying fMet-tRNA\sup{fMet} a the P
  site, and thus contains a  deacylated-tRNA\sup{fMet} at the P site. }

{\tempf $^\S$ 1.0 $\mu M$ and 0.5 $\mu M$ EF-G in the presence of 1 mM
  GDPNP (a non-hydrolyzable GTP analog)  were added to PMN\sub{fMet},
  respectively. }

\setcounter{figure}{0}
\renewcommand{\thefigure}{\arabic{figure}}

\clearpage
\begin{figure}
  \begin{center}
    \begin{minipage}{7in}
      \includegraphics{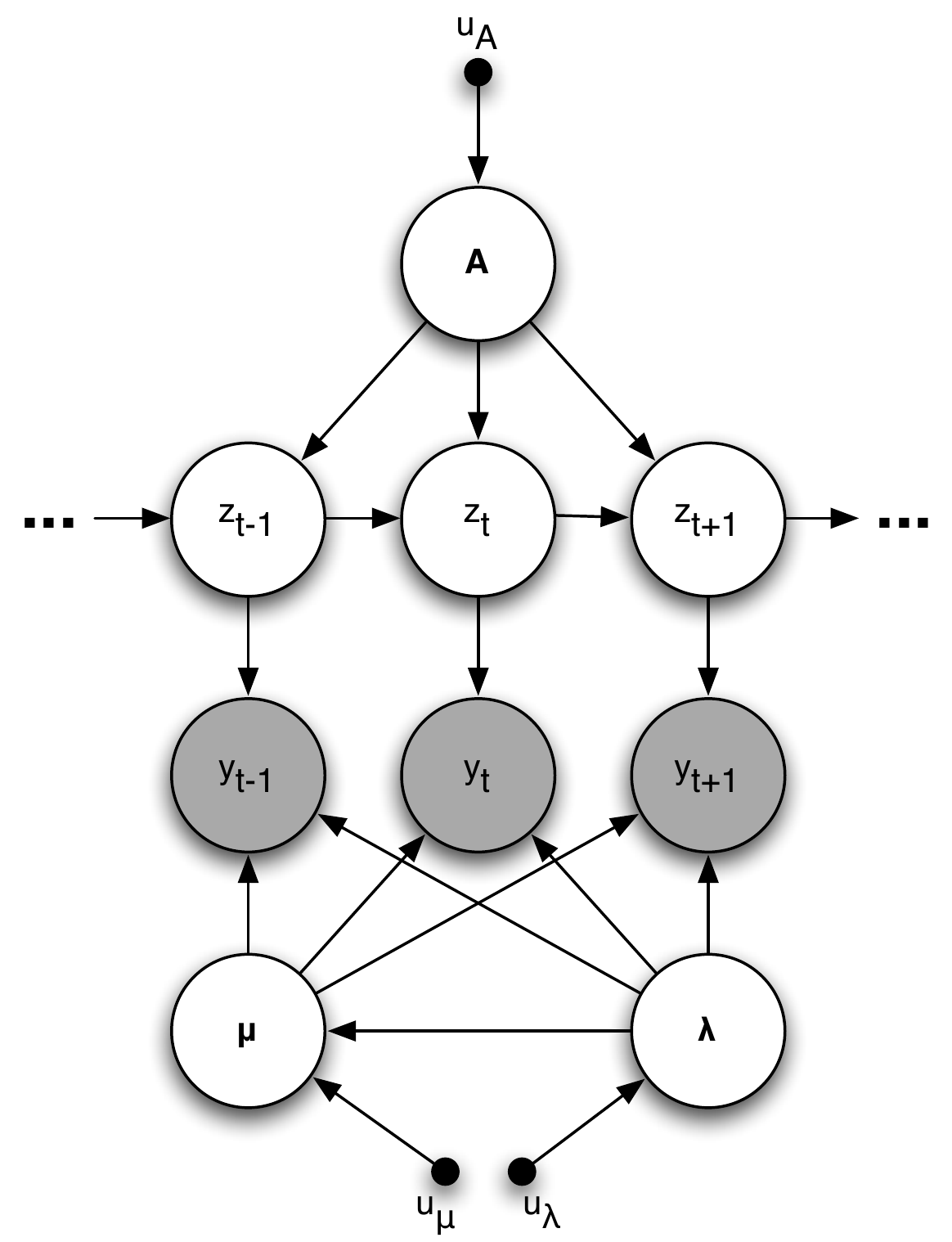}
      \caption{Graphical model representation of the HMM,
        corresponding to the factorization of the probability
        distribution in Eq. \ref{eq:hmm}. Each vertical slice
        represents a time slice $t=1,\ldots,T$, for which there is an
        observed FRET ratio $y_t$, given a hidden conformational state
        $z_t \in 1,\ldots,K$. Transitions between conformational
        states are represented by the dependencies between $z_t$ and
        $z_{t-1}$. Parameters are also shown as random variables, with
        arrows indicating the dependence of the observed and hidden
        variables. Parameters for the probability distributions over
        parameters (Sec.~\ref{sec:hyperparameters}) are shown as solid black
        circles. }
      \label{fig:graphicalmodel}
    \end{minipage}
  \end{center}
\end{figure}

\begin{figure}
  \begin{center}
    \begin{minipage}{7in}
      \includegraphics{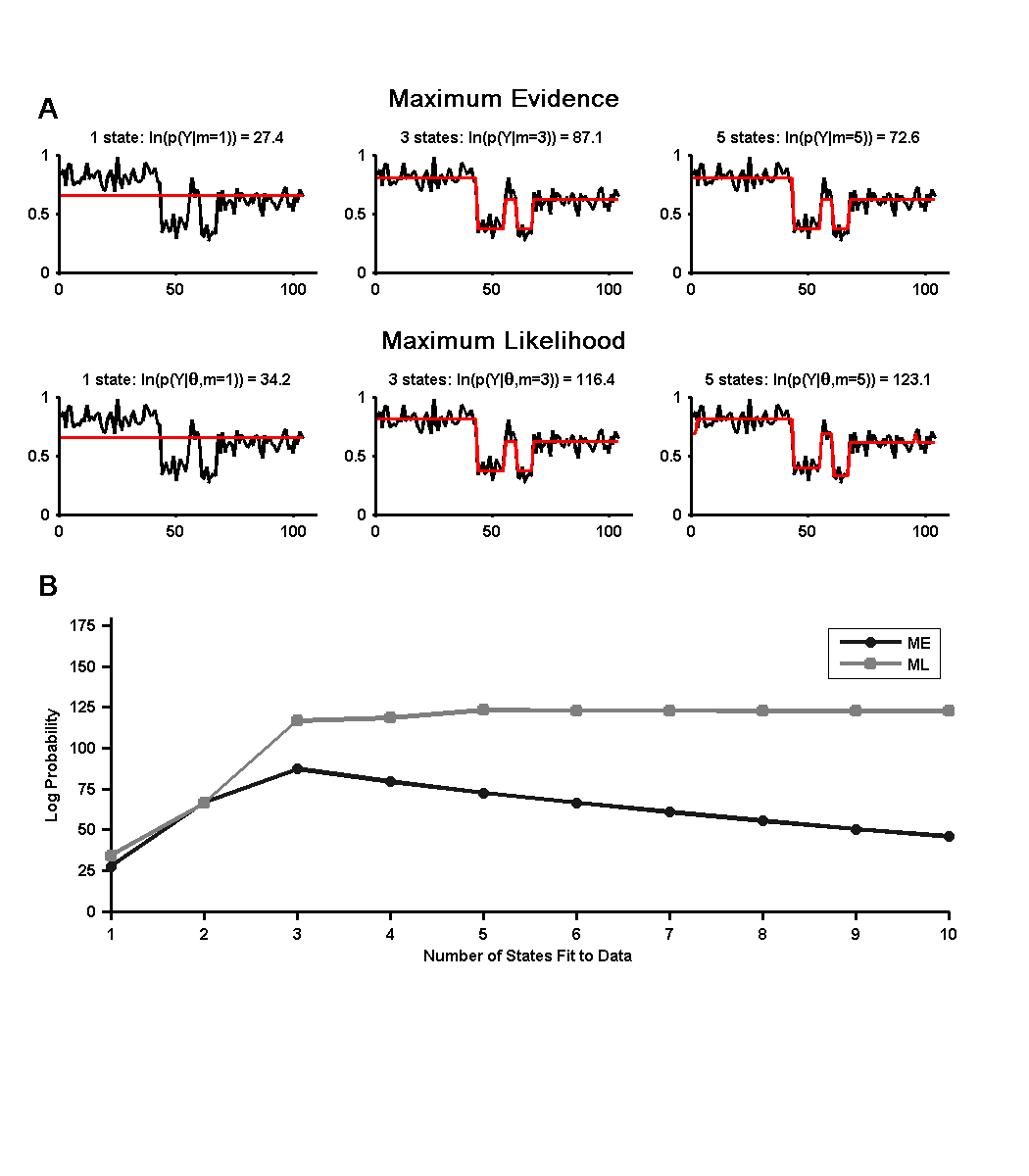}
      \vspace{-1in}
      \caption{ A single (synthetic) FRET trace analyzed by \vbf\ and
        \mlm. The trace contains 3 hidden states. A) (Top) idealize
        traces inferred by \vbf\ when $K = 1, K = 3$, and $K = 5$, as well as
        the corresponding log(evidence) for the inference. The data
        are under resolved when $K =1$, but for both $K = 3$ and $K = 5$ the
        correct number of states are populated.  (Bottom) idealized
        traces inferred by \mlm\ when $K = 1, K = 3$, and $K = 5$, as well as
        the corresponding log(likelihood). Inference when $K = 1$ and $K
        = 3$ are the same as for \vbf\ but the data are overfit when $K =
        5$. B) The log evidence from \vbf\ (black) and log likelihood from
        \mlm\ (gray) for $1 \leq K \leq 10$. The evidence is
        correctly maximized for $K = 3$, but the likelihood increases
        monotonically.}
      \label{fig:overfit}
    \end{minipage}
  \end{center}
\end{figure}

\clearpage
\begin{figure}
  \begin{center}
    \includegraphics{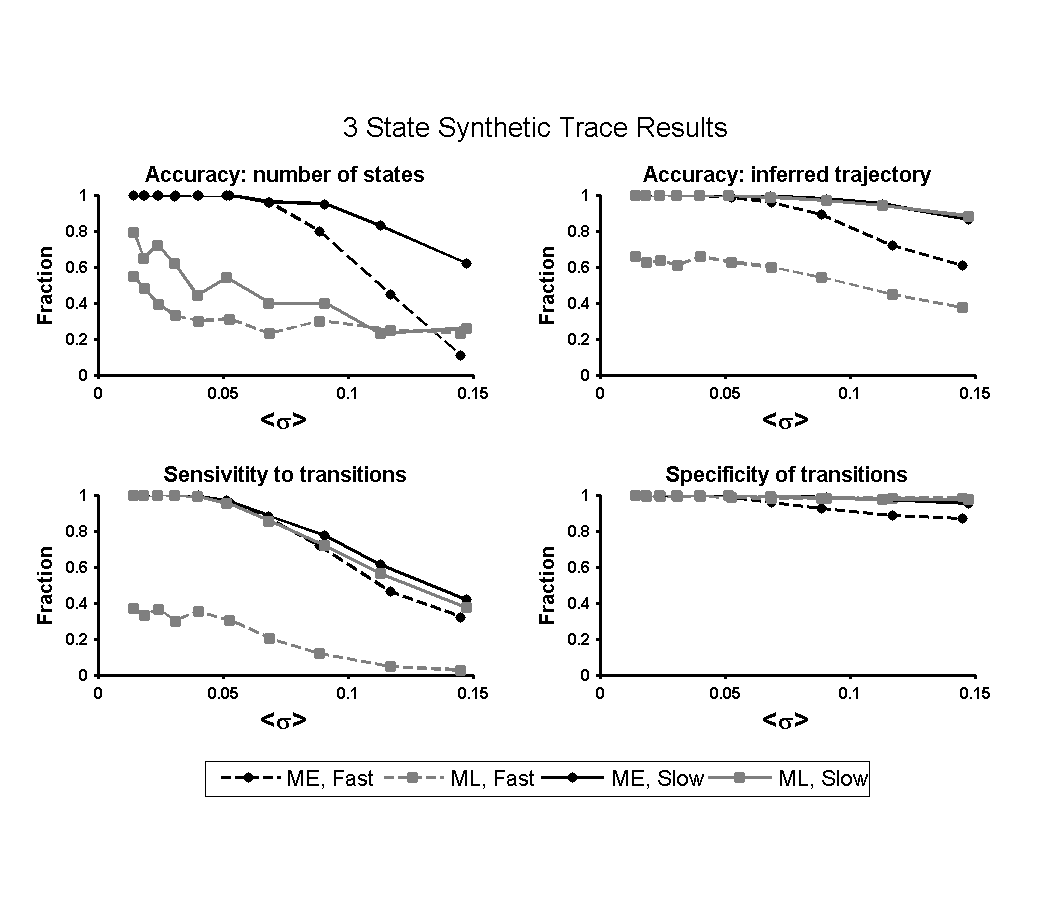}
    \caption{Comparison of \vbf\ and \mlm\ as a function of increasing
      hidden state noise. Fast transitioning (hidden state mean
      lifetime of 4 time steps) and slow transitioning (hidden state
      mean lifetime of 15 time steps) traces were created and analyzed
      separately. Each data point represents the average value taken
      over 100 traces. (Top left) $p(|\bzhat|=|\bz_0|)$: the
      probability in any trace of inferring the correct number of
      states.  (Top right) $p(\bzhat=\bz_0)$: the probability in any
      trace at any time of inferring the correct state. (Bottom left)
      sensitivity to true transitions: the fraction of time the
      correct FRET state was inferred during FRET trajectories.
      (Bottom right)  specificity of inferred transitions: the
      probability in any trace at any time that the true trace $\bz_0$
      does not exhibit a transition, given that $\bzhat$ does not.
      Error bars on all plots were omitted for clarity and because the
      data plotted represent mean success rates for Bernoulli
      processes (and, therefore, determine the variances of the data
      as well).}
      \label{fig:gauss3results}
  \end{center}
\end{figure}

\clearpage
\begin{figure}
  \begin{center}
    \includegraphics[width=1.0\textwidth]{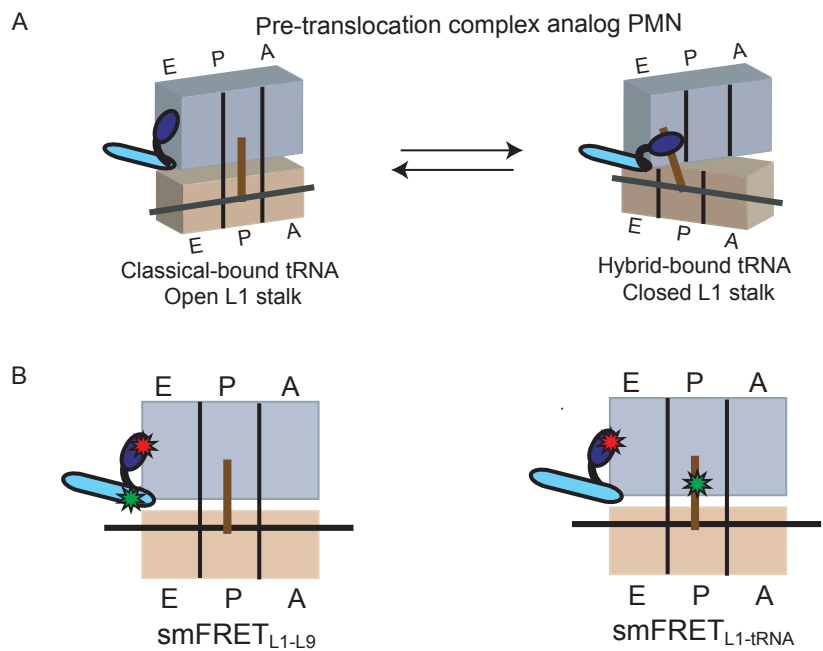}
    \vspace{-7.5cm} \caption{Conformational rearrangements within
        a pre-translocation (PMN) complex and smFRET labeling schemes.
        (A) Cartoon representation of a PMN complex analog. The small
      and large ribosomal subunits are shown in tan and
      lavender, respectively, with the L1 stalk depicted in dark blue,
      and ribosomal protein L9 in cyan. The aminoacyl-, peptidyl- and
      deacylated-tRNA binding sites are labeled as A, P and E,
      respectively, and the P-site tRNA is depicted as a brown
      line. PMN complex analogs are generated by adding the antibiotic
      puromycin to a post-translocation complex carrying a
      deacylated-tRNA at the E site and a peptidyl-tRNA at the P
      site.
      The resulting PMN complex analog exists in a thermally-driven
      dynamic equilibrium between two major conformational states in
      which the P-site tRNA fluctuations between classical and hybrid
      configurations correlate with the L1 stalk fluctuations between
      open and closed conformations. (B) Two labeling schemes have
      been developed in order to investigate PMN complex dynamics
      using smFRET. PMN complexes are cartooned as in (A) with Cy3 and
      Cy5 depicted as green and red stars,
      respectively. smFRET\sub{L1-L9} (left), which involves a Cy5
      label on ribosomal protein L1 within the L1 stalk and a Cy3
      label on ribosomal protein L9 at the base of the L1 stalk,
      reports on the intrinsic conformational dynamics of the L1
      stalk. smFRET\sub{L1-tRNA} (right), which involves a Cy5 label
      on ribosomal protein within the L1 stalk as in smFRET\sub{L1-L9}
      and a Cy3 label on the P-site tRNA, reports on the formation and
      disruption of a direct interaction between the closed L1 stalk
      and the hybrid bound P-site tRNA.}
    \label{fig:ribosome}
  \end{center}
\end{figure}

\clearpage
\begin{figure}
  \begin{center}
    \includegraphics[width=0.9\textwidth]{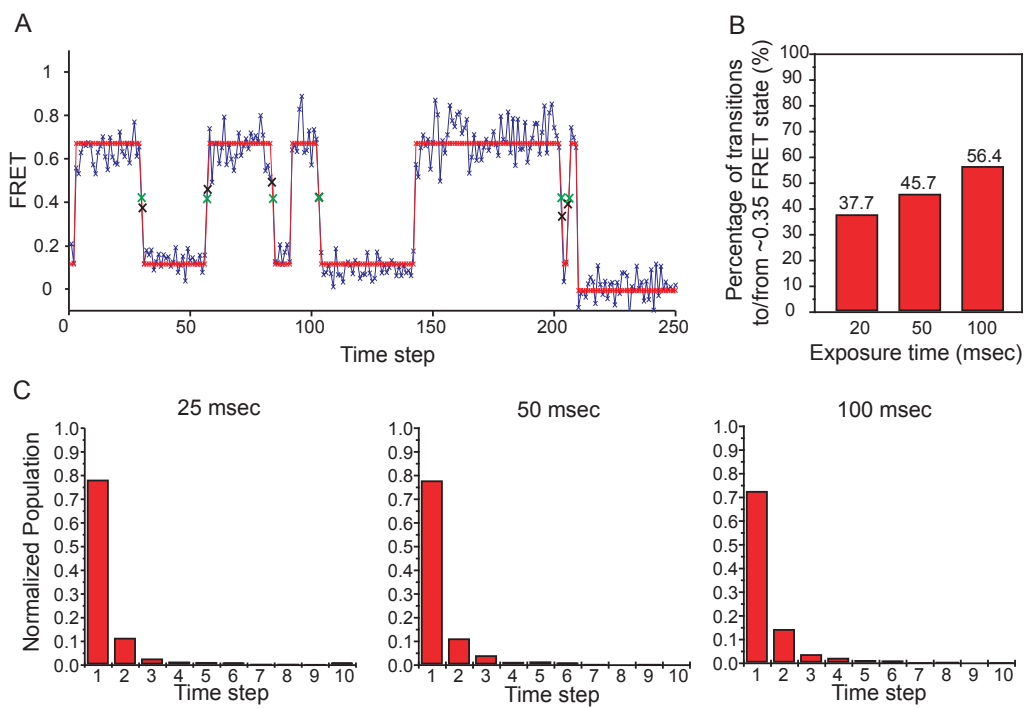}
    \vspace{-7cm} \caption{ Analysis of the smFRET\sub{L1-tRNA}
       $\fmid$ state.  A) A representative
      smFRET\sub{L1-tRNA} trace idealized by \vbf, taken from the 50 msec exposure
      time data set. Both the observed data (blue) and idealized path
      (red) are shown. Individual data points, real and idealized, are
      shown as Xs. To emphasize the data at or near $\fmid$,
      the Xs are enlarged and the observed and idealized data are show
      in black and green, respectively. (B) Bar graph of the
      percentages of transitions to or from the $\fmid$ state  under 25
      msec, 50 msec and 100 msec CCD integration time. (C) Normalized
      population histograms of dwell time spent at the $\fmid$
      state under 25 msec, 50 msec, and 100 msec CCD integration time.}

    \label{fig:blur}
  \end{center}
\end{figure}

\clearpage

\appendix
\section{$\chi^2$ and maximum likelihood}
\label{sec:chisquared}
	Minimization of squared loss is most commonly derived in the
	natural sciences by asserting that `error', the difference between
	parameterized model prediction and experimental data, is additive,
	 normally distributed, and independent for each example (here
	indexed by $i$):
	\beq
	y_i=f_\theta(x_i)+\xi_i; ~~ \xi_i\sim \mathcal{N}(\xi| 0,\sigma).
	\eeq
	
	This notation emphasizes that the model $f$ depends on parameters
	$\theta$, and the $\sim$ indicates the distribution from which
	the error $\xi_i$ on the $i^{\rm th}$ observation is drawn (i.e.,
	the Gaussian or normal distribution 
	and variance $\sigma$). Assuming independent and identically
	distributed observations, the probability of all the $N$ data
	${\bf y}=\{y_i\}_{i=1}^{i=N}$ is then the {\it
	likelihood} \beq L=p({\bf
	y}|\theta)=\Pi_{i=1}^{i=N} \mathcal{N}(y_i-f_\theta(x_i)|0,\sigma)= \frac{\re{-\chi^2}}{(\sqrt{2\pi \sigma})^N} \eeq
	with the usual $\chi^2=\sum_{i=1}^{i=N} (y_i-f_\theta(x_i))^2/\sigma^2$
	arising as a linear term in the logarithm of the likelihood $\ell$:
	\beq
	\ell\equiv \ln L=-\chi^2+{\frac{N}{2}}\ln 2\pi\sigma.
\label{eqn:elldef}
	\eeq
	Minimization of $\chi^2$, is thus derived from the more general principle
	of 
	ML: the parameters $\theta_*$ chosen are those which
	are the most likely.

\section{``BIC'': an intuition-building heuristic}
	\label{sec:BIC}
	\label{sec:bic}
	
	Often, explicit calculation of $\pyGk$ is computationally difficult,
	and one resorts to approximation. 
	For example, if the likelihood 
$\pyGtk$
	is sharply and uniquely peaked as a function of $\btk$, meaning
	that there is one unique maximum, Schwartz \cite{schwarz1978edm} suggested
	a pair of approximations: (i) Taylor expansion of $\ell(\bt)$ (from Eq. \ref{eqn:elldef})
	and Laplace approximation of the integral; and (ii) replacing the
	second derivative of $\ell(\bt)$ by its asymptotic behavior in the
	limit $\{K,N\}\goto \infinity$. 
	The first approximation reads
	\beq
	\pyGk=\int \dbtk {\re{\ell(\bt)}}\ptGk\approx 
	\re{\ell_*}
	\ptsGk
	\sqrt{\left|\frac{2\pi}{H}\right|}
	\eeq
	where $\ell_*=\ell(\theta_*)$ is the ML
	over all parameters $\theta$, and 
	the $K\times K$ matrix $H$, also termed the Hessian,
	is the matrix of derivatives  (evaluated at $\ts$)
	\beq
	H_{\alpha\beta}\equiv
	\frac{\partial^2 \ell(\bf\theta)}{\partial\theta_\alpha\partial \theta_\beta}.
	\eeq
	In the case of independent data the derivative of $\ell$ 
	is a sum of $N$ independent terms, and the determinant of the 
	Hessian scales as
	$N^K$ in the limit of infinite data $N$ and infinitely
	many $K$ equally-important parameters $\theta_\alpha$. Under
	this pair of asymptotic approximations, then,
	\beq
	\pyGk\approx 
	\re{\ell_*}
	\ptsGk 
	\sqrt{\left|\frac{2\pi}{H}\right|}\approx C(K,N)\re{(\ell_*-K/2\ln N)}.
	\eeq
	The exponent is sometimes referred to as the {\it Bayesian Information
	Criterion} or BIC; 
for clarity it is worth noting, though,
 	that it does not depend on the prior (the
 	most common meaning of the adjective `Bayesian' 
	in statistics) and
	that it is derived without any appeal to or use of 
	information theory. The usage of such an algebraic expression
	alone, ignoring the possible dependence of terms lumped into $C(K,N)$ 
	(i.e., treating $C(K,N)$ as a constant) is a simple\footnote{
	Note that, although use of the BIC obviates determining
	many facets of one's model and its relation to the
	data, we still need to know the error bars $\sigma$, which
	appear in $\ell$.},
	intuitive,
	and appealing approach to model selection. The increase in $\ell_*$ 
	as $K$ increases is penalized by the term $-K/2\ln N$, selecting the
	optimal model indexed by $K_*$, the maximizer of the BIC.

	In the case of FRET data
	the likelihood is complicated by the presence of a hidden
	 state $z_i$ (the discrete conformational state 
	of the molecule which gives rise to the observed FRET 
ratio),
	meaning that
	the evidence $\pyGk$ has the richer formulation (suppressing
	the cluttering superscripts $K$ on the hidden and manifest variables
	$z$ and $\theta$, respectively)
	\beq
	\pyGk=\sum_{\bf z}\int \dbtk \pyzGtk \pzGtk \ptGk.
	\eeq
	This rich structure renders completely inappropriate the
	assumptions of the BIC derivation above: among other
	problems, the hidden variables
	will be modeled by a Markovian dynamic, 
	 coupling each
	of the example data
	(and thus violating the assumption of $N$ independent data); and 
	the permutation symmetry of the labels on these
	violates the assumption that the likelihood is sharply and singly peaked --
        rather there are $K!$ such peaks from the possible re-labelings of the
        states.

\section{Proof of variational relation}
\label{sec:dklproof}

We provide a proof of the variational relation in
Eq. \ref{eq:DKL}. We start with the desired quantity, the evidence
$\pyGk$, and multiply by one,
\beq
  \ln \pyGk &=& \left[ \sum_{\z} \int d\theta \qzt \right] \ln \pyGk,
\eeq
valid for any normalized probability distribution $\qzt$. 
We then use the definition of conditional probability to write
\beq
  \pyztGk = \pztGyk \pyGk.
\eeq
We use this to rewrite the argument of the logarithm and multiply by
one yet again:
\beq
\ln \pyGk &=& \sum_{\z} \int d\theta \qzt \ln \left[{\pyztGk \over \pztGyk}\right] \\
          &=& \sum_{\z} \int d\theta \qzt \ln \left[{\pyztGk \qzt \over \qzt \pztGyk}\right] \\
          &=& \sum_{\z} \int d\theta \qzt \ln \left[{\pyztGk \over \qzt}\right] +
              \sum_{\z} \int d\theta \qzt \ln \left[{\qzt \over \pztGyk}\right],
\eeq
where in the last line we have separated logarithm to decompose the
integral into two parts. We recognize the rightmost term as the
Kullback-Leibler divergence between $\qzt$ and $\pztGyk$,
\beq
  D_{KL}(\qzt||\pztGyk) &=& \sum_{\z} \int d\theta \qzt \ln \left[{\qzt \over \pztGyk}\right]
\eeq
and define the remaining term as the negative of the {\it free energy},
\beq
  F[\qzt] = -\sum_{\z} \int d\theta \qzt \ln \left[{\pyztGk \over \qzt}\right],
  \label{eq:F}
\eeq
which results in the variational relation presented in
Eq. \ref{eq:DKL},
\beq
  \ln \pyGk &=& -F[\qzt] + D_{KL}(\qzt||\pztGyk).
  \label{eq:DKL2}
\eeq
This completes the proof of the variational relation and offers
several insights.

The first is that the free energy is strictly bounded by the
log-evidence, as the Kullback-Leibler (KL) divergence is a
non-negative quantity, proven through an application of Jensen's
inequality \cite{Jensen:1906} (an extension of the definition of
convexity). Thus we have reduced the problem of approximating the
evidence to that of finding the distribution $\qzt$ which is
``closest'' to the true (and intractable) posterior $\pztGyk$ in the
KL sense. As per Eq. \ref{eq:DKL2}, we see that this is equivalent to
minimizing the free energy $F[\qzt]$ as a functional of $\qzt$.  This
observation motivates the variational Bayes Expectation Maximization
algorithm, in which a specific factorization for $\qzt$ is chosen as
to make calculation of $F[\qzt]$ tractable, and iterative coordinate
ascent is performed to find a local minimum.

In addition, we provide motivation for the term ``free energy'',
rewriting Eq. \ref{eq:F} by decomposing the logarithm:
\beq
  F[\qzt] &=& -\sum_{\z} \int d\theta \qzt \ln \left[{\pyztGk \over \qzt}\right] \\
          &=& -\sum_{\z} \int d\theta \qzt \ln \pyztGk 
              +\sum_{\z} \int d\theta \qzt \ln \qzt.
\eeq
Recognizing the negative log-probability in the first term as an
energy (as in the Boltzmann distribution) and the second term as the
information entropy \cite{Shannon:1948,Shannon:1948a} of $\qzt$, i.e.
\beq
  E &\equiv& -\ln \pyztGk \\
  S &\equiv& -\sum_{\z} \int d\theta \qzt \ln \qzt.
\eeq
Thus we can rewrite \ref{eq:F} as
\beq
  F = \<E\>-TS,
\eeq
(with unit ``temperature'' T) where the angled brackets denote
expectation under the variational distribution $\qzt$. This familiar
form from statistical physics offers the following interpretation: in
approximating the evidence (and posterior), we seek to minimize the
free energy by finding a distribution $\qzt$ that balances minimizing
the energy and maximizing entropy.

We 
encourage
the 
reader to 
enjoy the
texts \cite{MacKay:2003} and \cite{Bishop2006} for more 
pedagogical
discussions of variational methods.

\bibliographystyle{biophysj}
\bibliography{bronson}
\clearpage
\setcounter{page}{1}

\clearpage

\renewcommand{\thefigure}{S.\arabic{figure}}
\setcounter{figure}{0}
\renewcommand{\thetable}{S.\arabic{table}}
\setcounter{table}{0}

\setcounter{section}{18}
\section{Supporting material}
\label{sec:supporting}

\subsection{Methods}
\label{sec:methods}
\subsubsection{ML inference settings}
\label{sec:ML_setting}

Following the HaMMy user manual, ML
 analyses use $K_{max}+2$ states, where $K_{max}$ is either $K_0$ (the
 true number of states) in the case of synthetic data or simply 3 in
 the case of experimental data, as 1D FRET histograms suggest 
 two biophysical states and one photophysical state: the
 photobleached state. No additional complexity control was applied to
 the resulting parameters inferred from individual traces.  The default
 guess for the initial distribution of the means $\mu_z$ was used,
 i.e., uniform spacing between $0$ and $1$ \FRET.

 Also consistent with default settings, we use the parameters inferred
 using only one set of initial parameter-guesses.  Note that this
 differs from the usual implementation of expectation-maximization as
 a technique for performing ML (\cf
 \cite{Bishop2006}). Expectation-maximization (the maximization
 technique used in both HaMMy and \vbFRET) provably converges to a
 local optimum, and therefore the maximization typically is performed
 using many random restarts for parameter values. One possible reason
 to avoid this procedure is the inescapable pathology of ML for
 real-valued emissions (\eg in FRET data) and for which the width of
 each state is an inferred parameter: the optimization is ill-posed
 since the case in which one observation is assigned to a state of 0
 uncertainty is infinitely likely (\cf \cite{Bishop2006} Ch. 9:
 ``These singularities provide another example of the severe
 overfitting that can occur in a maximum likelihood approach.  We
 shall see that this difficulty does not occur if we adopt a Bayesian
 approach.'').
\label{sec:ml-restarts}

\subsubsection{ME inference settings}
\label{sec:ME_settings}
In analyzing synthetic and experimental data with \vbf, we
attempt each choice of 
$K=1,2,\ldots,K_{\rm max}+2$
with $K_{\rm max}$ as above. For synthetic data, 25 
random initial guesses were used for each of the traces; 
for experimental data, 100 initializations were used (though,
in our experience, little or no change in the optimization
was found after 25 initializations). 
As with all local optimization techniques, including expectation maximization
in ML or in ME, we use the parameters which give the optimum over all restarts
(here, the set of parameters specifying the approximating distribution $q$ which
gives the maximum evidence $\pyGk$).

\subsubsection{Rate constant calculations}
\label{sec:rates_methods}
Rates for the smFRET\sub{L1-L9} experimental data, both for \vbf\ and \mlm\ analyses,
were extracted as previously described
\citep{Gonzalez2008,Sam09}. 
First, the set of all idealized traces over all times is histogrammed
into 50 bins, evenly spaced between $-0.2$ and $1.2$ \FRET. The counts
in the resulting histogram are given to Origin 7.0, which learns a
Gaussian mixture model via expectation-maximization, using
user-supplied initial guesses for the three means (we used
$\mu=(0,0.35,0.55)$ \FRET).  Origin returns true means and variances
for each of the 3 states.  From these variances the width at half-max
for each mixture is determined, defining three acceptable ranges of
fret values.  (For this experiment, these ranges had widths of
approximately $.05$ \FRET.  We next re-scan the idealized traces and,
for each transition from one acceptable range to another, record the
dwell time (the total time spent within the range; any number of
inferred transition within one accepted range are ignored, effectively
smoothing of overfit idealized traces).  The cumulative distribution
of dwell times from a given state is now given to Origin 7.0 to infer
the most likely parameters, asserting exponential decay. The inverse of
the inferred time constant is the rate constant reported for that state.

\subsubsection{Generating synthetic data}
\label{sec:gen_synt}
Synthetic data were generated in MATLAB. Rather than 
testing the inference on data generated precisely by the
emissions model (one in which the scalar FRET signal is
taken to be normally-distributed in each state), we challenge
the inference by using a slightly more realistic distribution: one
that is normally-distributed in each of the two florophore colors.
That is, each synthetic trace was
created from a hidden Markov model with 2D Gaussian output (representing
the two florophore colors). The 2D
data $\bx_1,\bx_2$ were then FRET transformed using $\bff=\bx_2  /
(\bx_1+\bx_2)$; 
points such that $f\notin (0,1)$ were discarded.  

The 2D Gaussians are chosen so that, in any state $z$,
the sum of the means is 1000 ($\mu_z^1+\mu_z^2=1000\ \forall z$),
roughly corresponding to our experimental data. 
Variances were drawn from a uniform distribution
centered at each dimension's mean over a range given by 10\% of the mean.
The two components were allowed a nonzero covariance, also drawn
from a uniform distribution centered at 0, with a range given by
$1/2$ the smaller of the two means. We emphasize that these
choices are intended both to be consistent with the smFRET\sub{L1-L9}
and smFRET\sub{L1-tRNA}
data and {\it not} to match the algebraic expressions in the priors
used below, which would be a less challenging inference task (model
specification identically matching the generative process).

Increasingly noisy traces were generated by multiplying the 
covariance matrix of each hidden state by a constant. Ten 
constants, chosen log-linearly between 1 and 100, were used. 
The mean standard deviation of the FRET state noise in the 
resulting 1D traces varied from, approximately, $0.02< \sigma < 1.4$.

\subsection{Priors}
\label{sec:priors}

\subsubsection{Mathematical expressions for priors }
\label{sec:prior_math}

To calculate the model evidence, we treat the components of $\bt$ as
random variables. The vector $\vec{\pi}$ and each row of A are modeled
as Dirichlet distributions:
\beq
p(\vec{\pi}) &=& 
	\frac{\Gamma(\sum_{k=1}^{K}\upik)}{\prod_{k=1}^{K}\Gamma(\upik)} 
	\prod_{k=1}^{K}\pi_{k}^{\upik-1}
\\
p(a_{j1},...,a_{jK}) &=& 
	\frac{\Gamma(\sum_{k=1}^{K}\uajk)}{\prod_{k=1}^{K}\Gamma(\uajk)}
	\prod_{k=1}^{K}a_{jk}^{\uajk-1}
\eeq
The probabilities for each pair of $\mu_k$ and $\lambda_k$ are modeled
jointly as a Gaussian-Gamma distribution:
\beq
p(\mu_k, \lambda_k) = 
\sqrt{\frac{\ubk\lambda_k}{2\pi}}e^{-\oh\ubk\lambda_k(\mu_k
  - \umk)^2}\frac{1}{ \Gamma
  (\uvk/2)}(2\uwk)^{-\uvk/2}\lambda_k^{(\uvk/2)-1}e^{-\frac{\lambda_k}{2\uwk}}.
\eeq
The terms 
$\bupi$,
$\bua$,
$\bub$,
$\bum$,
$\buv$,
and 
$\buW$
are  called the {\it hyperparameters} for
the probability distributions over 
$\bt$.

\subsubsection{Hyperparameter settings}
\label{sec:hyperparameters}
Hyperparameters for \vbFRET\ were set so as to give
distributions consistent with experimental data and to
influence the inference as weakly as possible:
$u_{\pi}^k = 1$, $u_a^{jk} = 1$, $u_{\beta}^k
= 0.25$, $u_m^k = 0.5$, $u_v^k = 5$ and $u_W^k = 50$, for all values
of k.  
Qualitatively,
these hyperparameters priors correspond to probability distributions
over the hidden states such that it is most probable that the hidden
states are equally likely to be occupied and equally likely to be
transitioned to. 
Quantitatively, they yield  $\<\mu_k\>= 0.5$ and typical $\sigma\approx
0.08$, consistent with experimental observation.
($1/\sqrt{{\rm mode}(\lambda_k)}=1/\sqrt{150}\approx 0.08 ~\forall k$).

\subsubsection{Sensitivity to hyperparameter settings}
\label{sec:hpar_results}
\begin{figure}
   \begin{center}
      \includegraphics{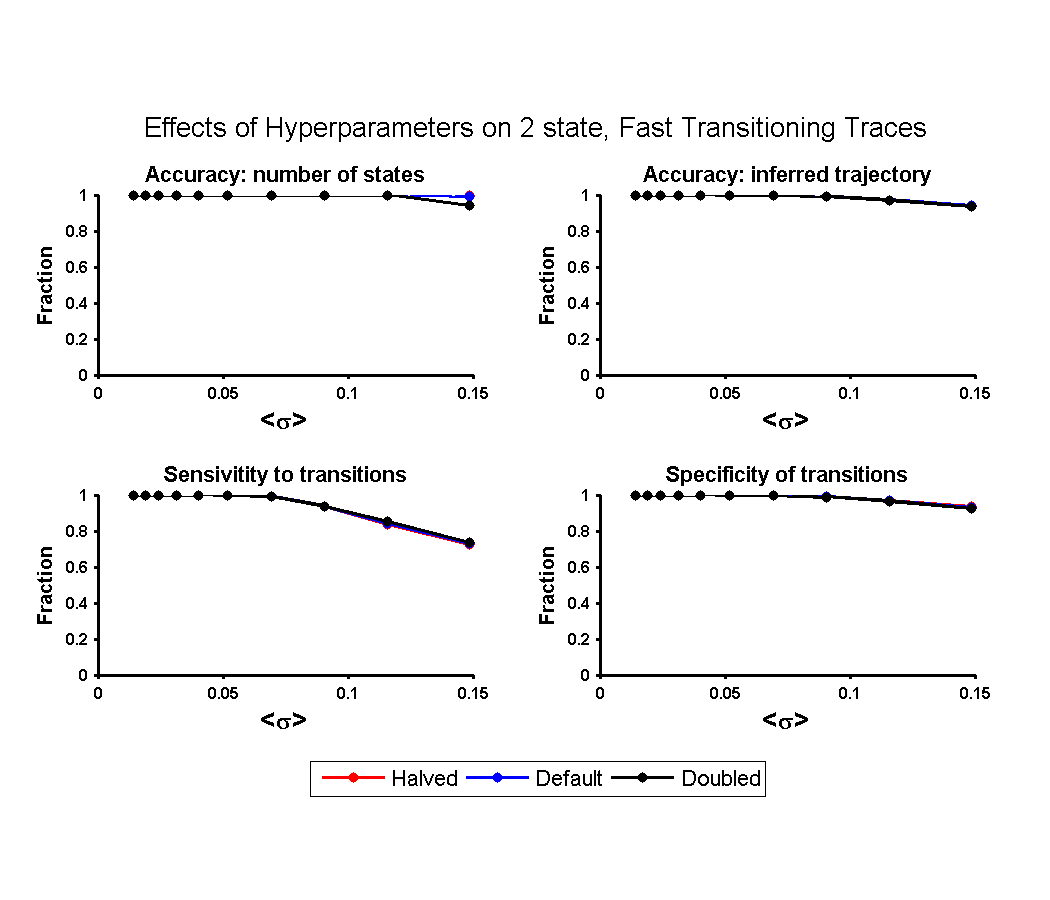}
      \caption{Effects of hyperparameter settings on
        fast-transitioning, two state traces.}
      \label{fig:hpar_synthetic_2fast}
   \end{center}
\end{figure}

\begin{figure}
   \begin{center}
      \includegraphics{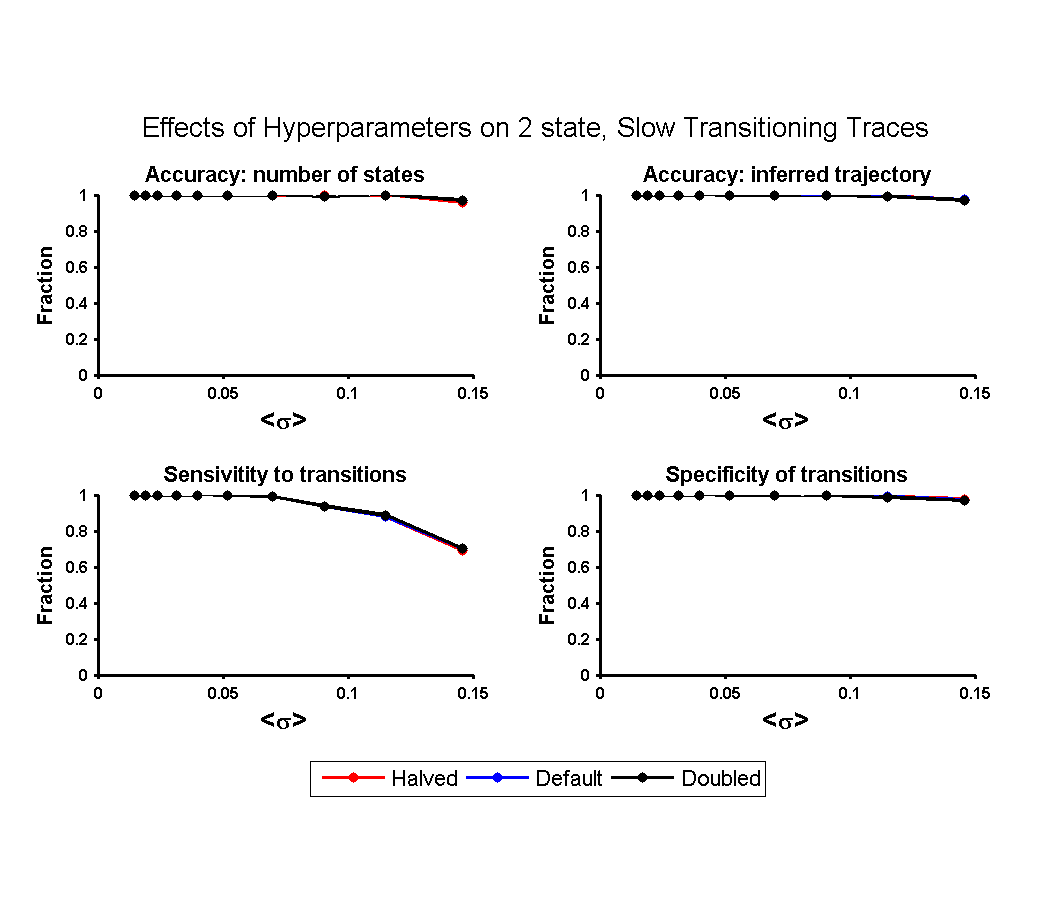}
      \caption{Effects of hyperparameter settings on
        slow-transitioning, two state traces.}
      \label{fig:hpar_synthetic_2slow}
   \end{center}
\end{figure}

\begin{figure}
   \begin{center}
      \includegraphics{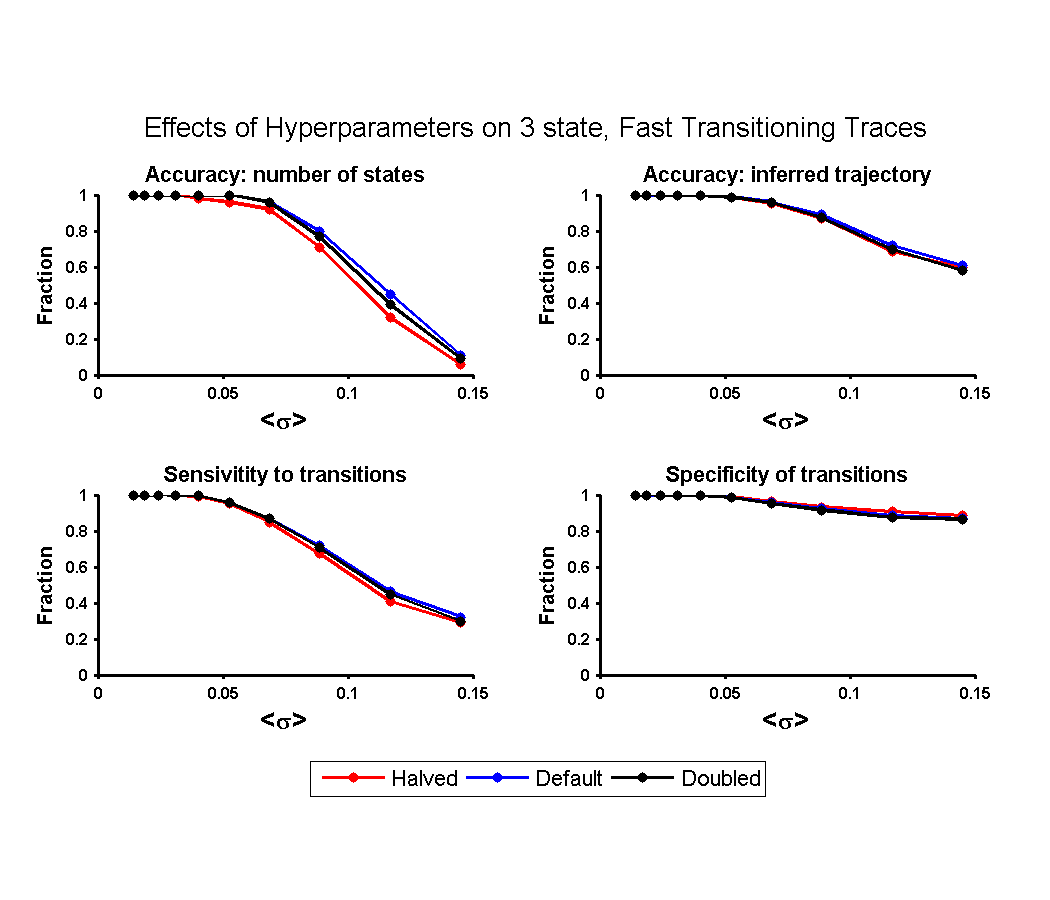}
      \caption{Effects of hyperparameter settings on
        fast-transitioning, three state traces.}
      \label{fig:hpar_synthetic_3fast}
   \end{center}
\end{figure}

\begin{figure}
   \begin{center}
      \includegraphics{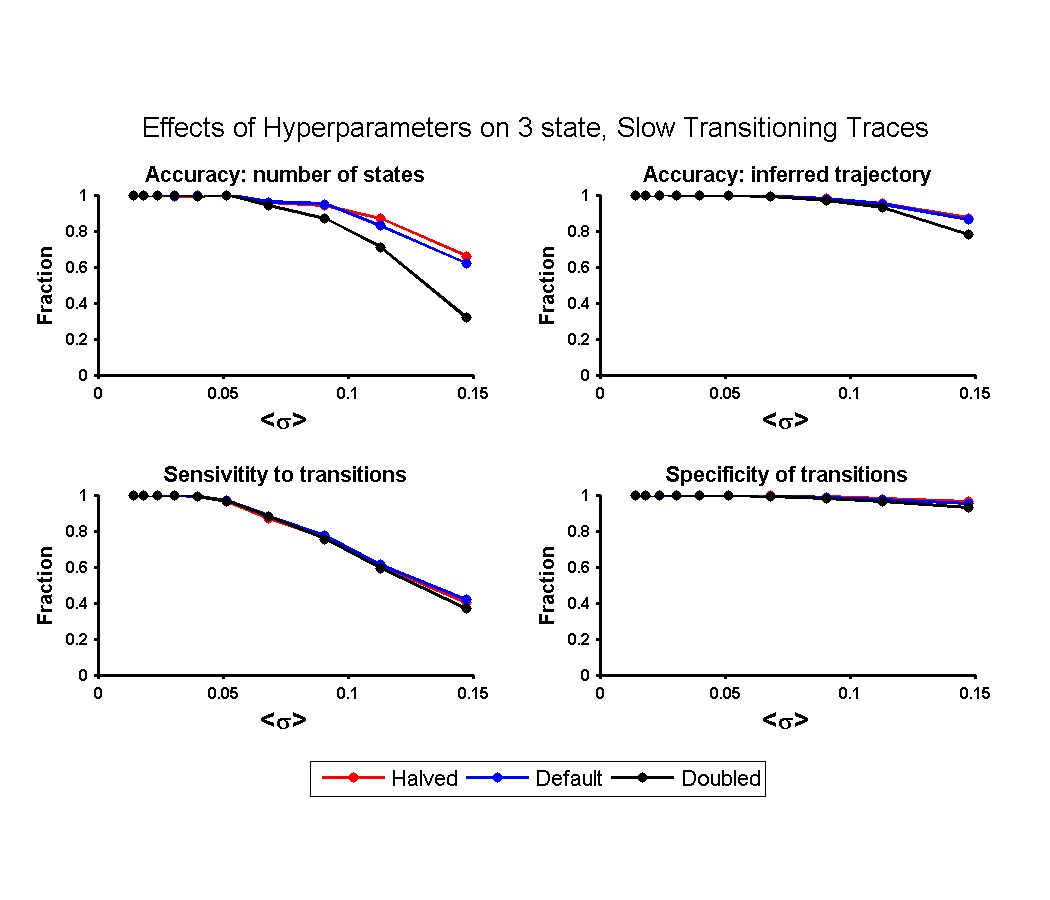}
      \caption{Effects of hyperparameter settings on
        slow-transitioning, three state traces.}
      \label{fig:hpar_synthetic_3slow}
   \end{center}
\end{figure}

\begin{figure}
  \begin{center}
    \includegraphics{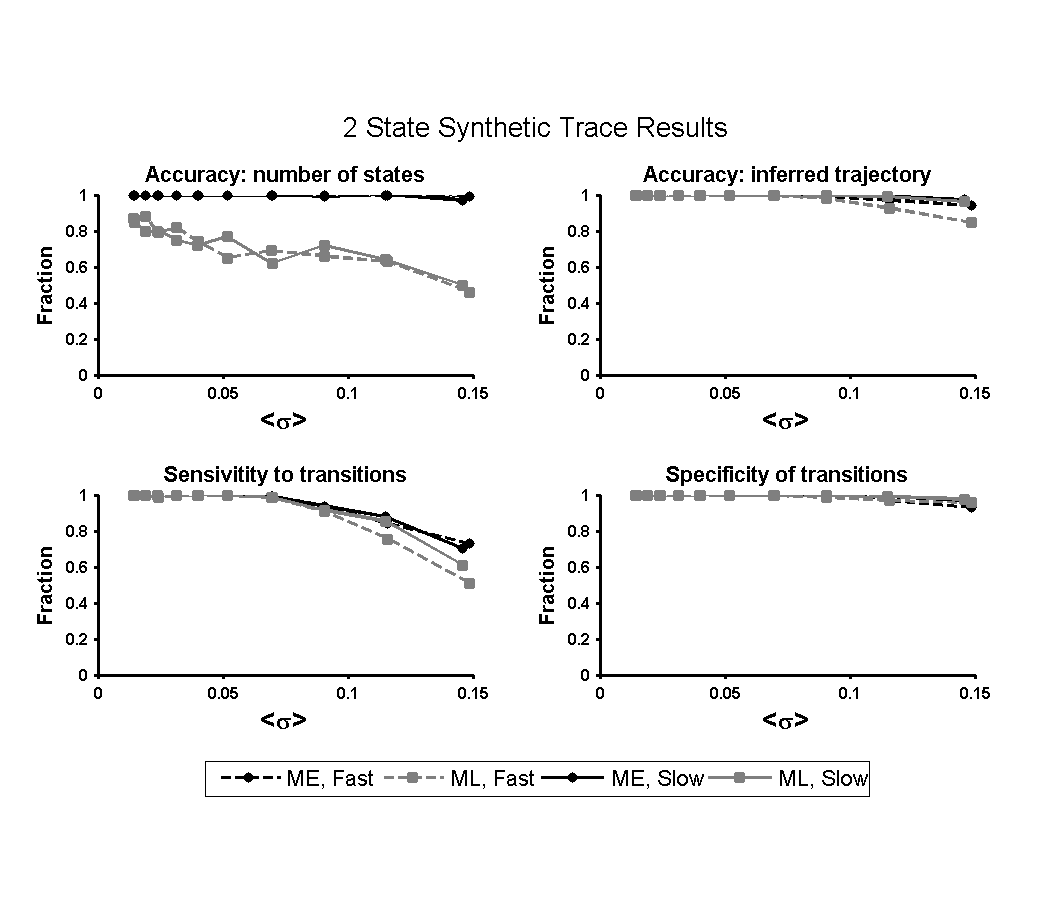}
     \caption{Synthetic results for two state traces.}
    \label{fig:synthetic_results2}
  \end{center}
\end{figure}

One standard approach 
\cite{McCulloch1991141,kass1995bf}
to sensitivity
analysis is to halve and double hyperparameters and recompute
the evidence for different models.
The sensitivity of \vbf\ inference on hyperparameter settings was
investigated on both experimental and synthetic data. First, the two and
three state traces from Fig.~\ref{fig:gauss3results} and
Fig.~\ref{fig:synthetic_results2} were reanalyzed with all the
hyperparameters set to one half their default values and twice their
default values (Figs.~\ref{fig:hpar_synthetic_2fast},
\ref{fig:hpar_synthetic_2slow}, \ref{fig:hpar_synthetic_3fast},
\ref{fig:hpar_synthetic_3slow}). One hyperparameter, the prior on the
mean of each Gaussian, was not changed during this analysis, since its
value is set to 0.5 based on a symmetry argument.

The results show a relative insensitivity to the hyperparameter values
over the settings considered. The largest difference in inference
accuracy between the different settings was for the noisy,
slow-transitioning traces shown in
Fig.~\ref{fig:hpar_synthetic_3slow}, when the hyperparameters were
doubled. Interestingly, these traces are harder to resolve than the
two state traces but not as difficult to resolve as the noisy,
fast-transitioning three state traces. A possible explanation for this
behavior is that the two state trace results are insensitive to
hyperparameter settings because the data are easy enough to resolve
and the noisy, fast-transitioning three state traces are insensitive
to hyper parameter settings because they are too hard to resolve. The
noisy, slow-transition states are on the border of being resolvable,
so using a prior that more closely matches the true parameters of the
model yields more accurate results.
Additionally, the three
state, slow-transition data has the highest probability of having a
sparsely populated state (i.e. one that is only present for a few time
steps in a trace). When $\sigma$ is large, these sparsely populated
states become harder to identify as distinct states, which may explain
why $p(|\hat{\bz}| = |\bz_0|)$ decreases more than $p(\hat{\bz} =
\bz_0)$, sensitivity or specificity .

\clearpage To investigate further the effects of the hyperparameter
settings on \vbf\ inference, the experimental data from
Table~\ref{table:real_dat} were reanalyzed using a more strongly
diagonal transition matrix prior (Table~\ref{table:real_dat_sup}). In
this second prior, the diagonal terms of the transition matrix were
set to 1 and the off-diagonal terms were set to 0.05, loosely
corresponding to a prior belief that the ribosome was 10x more likely
to remain in its current state than transition to a new one. For all
of the data, the transition rates calculated with both hyperparameter
settings are within error of each other for all transition rates.

\newcommand{\DEF}{Default}
\newcommand{\DIA}{Diagonal}

\begin{table}[h]
  
  \caption{Effect of hyperparameters on transition rate inference}
  \begin{tabular}{l l c c }
    & & & \\
    Data set$^*$ & Settings & \kclose & \kopen \\ \hline \hline

    &&& \\[-3.5mm]

    PMN\sub{Phe}$^{\dagger}$ & \DEF  &  $0.66 \pm 0.05$ & $1.0 \pm 0.2$ \\

    & \DIA & $0.66 \pm 0.04$ & $1.0 \pm 0.2$\\ \hline 

    &&& \\[-3.5mm]

    PMN\sub{fMet}$^{\ddagger}$ & \DEF &  $0.53 \pm 0.08$ & $1.7 \pm 0.3$ \\

    & \DIA  & $0.52 \pm 0.09$ & $1.7 \pm 0.1$ \\ \hline

    &&& \\[-3.5mm]

    PMN\sub{fMet+EFG}  & \DEF &  $3.1 \pm 0.6$ &
    $1.3 \pm 0.2$ \\

     ($1\ \mu M$)$^{\S}$ & \DIA & $2.8 \pm 0.5$ & $1.3 \pm 0.1$ \\\hline

    &&& \\[-3.5mm]

    PMN\sub{fMet+EFG}  & \DEF & $2.6 \pm 0.6$ & $1.5 \pm 0.1$ \\

     ($0.5\ \mu M$)$^{\S}$ & \DIA & $2.6 \pm 0.5$ & $1.4 \pm 0.1$ \\\hline

 \end{tabular}
 \label{table:real_dat_sup}
\end{table}

{\tempf $^*$ Rates reported here are the average and standard deviation
  from three or four independent data sets. Rates were not corrected
  for photobleaching of the fluorophores.}

{\tempf $^\dagger$ PMN\sub{Phe} was prepared by adding the antibiotic
  puromycin to a post-translocation complex carrying
  deacylated-tRNA\sup{fMet} at the E site and fMet-Phe-tRNA\sup{Phe}
  at the P site, and thus contains a deacylated-tRNA\sup{Phe} at the P
  site.}

{\tempf $^\ddagger$ PMN\sub{fMet} was prepared by adding the antibiotic
  puromycin to an initiation complex carrying fMet-tRNA\sup{fMet} a the P
  site, and thus contains a  deacylated-tRNA\sup{fMet} at the P site. }

{\tempf $^\S$ 1.0 $\mu M$ and 0.5 $\mu M$ EF-G in the presence of 1 mM
  GDPNP (a non-hydrolyzable GTP analog)  were added to PMN\sub{fMet},
  respectively. }

\subsection{Synthetic validation -- 2 and 4 state traces} 
\label{sec:validation_sup}
Synthetic data for 2 FRET state traces (fast- and slow-transitioning,
smFRET state means at 0.3 and 0.7 FRET) and 4 FRET state traces
(fast-transitioning only, smFRET state means at 0.21, 0.41, 0.61 and
0.81 FRET) were generated and analyzed exactly as the traces in
Fig.~2. The results are qualitatively similar to those in
Fig.~2. Inference accuracy begins to decrease at a lower noise level
as more FRET states are added to the traces. This should not be
surprising, though, since the states are more closely spaced as the
number of states increases, and therefore should be harder to resolve.

\begin{figure}
  \begin{center}
    \includegraphics{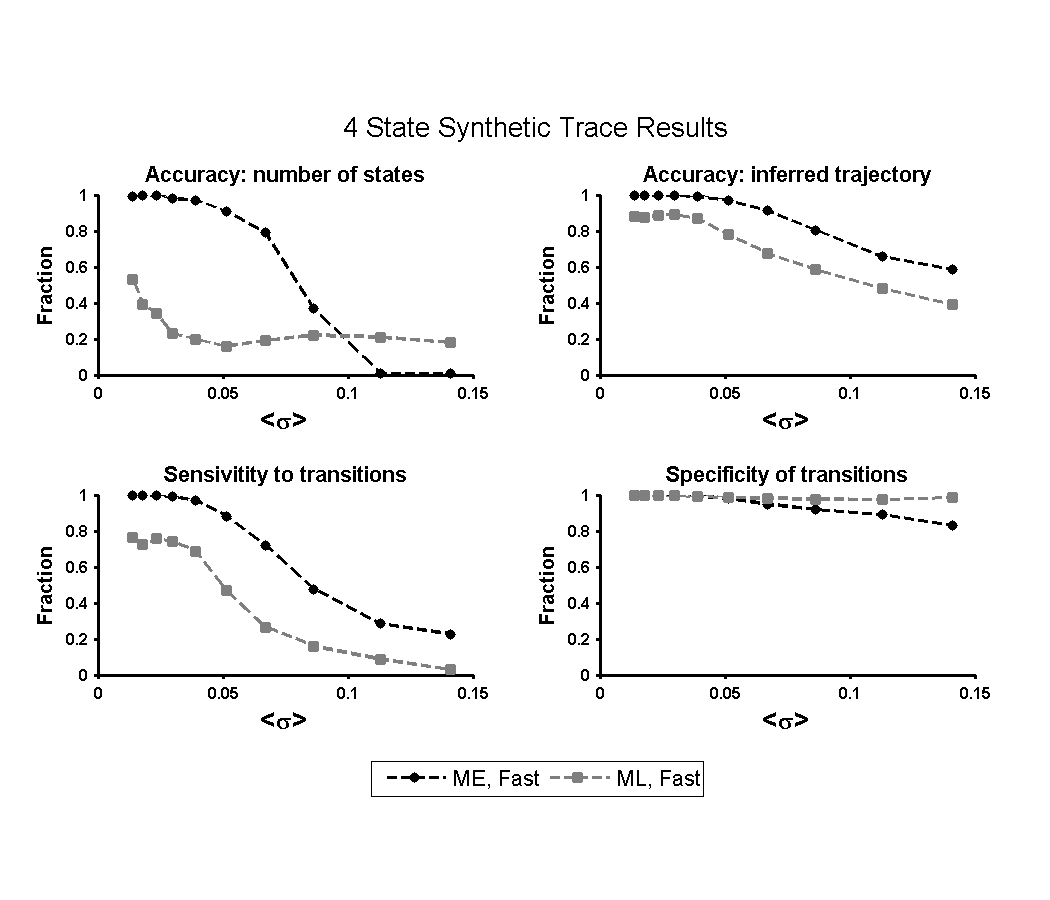}
     \caption{Synthetic results for four state traces.}
    \label{fig:synthetic_results4}
  \end{center}
\end{figure}


\subsection{Blur state TDPs}
\begin{figure}[h]
  \begin{center}
    \includegraphics[width=0.9\textwidth]{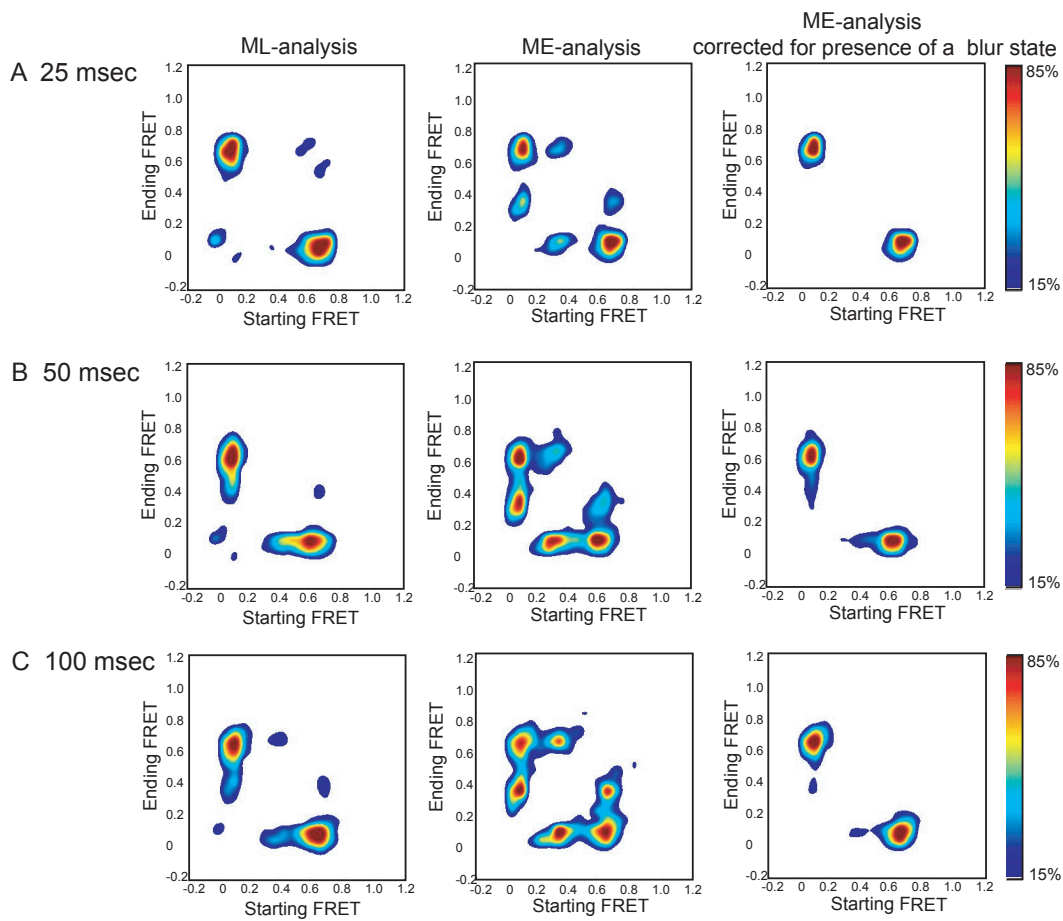}
    \vspace{-6cm}\caption{Transition density plots (TDP) of
      smFRET\sub{L1-tRNA} PMN\sub{fMet+EF-G} derived from \vbf\ and \mlm\
      analysis with different CCD integration times.  TDPs are contour
      plots showing the kernel density estimation of the transitions
      in idealized traces (with starting and ending FRET values of the
      transitions as the X and Y axes, respectively). Note that
      transitions to short-lived or nearby states count with equal
      weight as those to long-lived states in a TDP. This should
      not be confused with a time-density plot, which illustrates the
      probability of observing a pair of experimental values at two
      different times $p(y(t),y(t+\delta t))$, which can be made from
      the FRET data themselves without appealing to statistical
      inference. The plots show \mlm\ (left), \vbf\
      (middle) and \vbf\ analysis corrected for the presence of a blur
      state (right) 
      Contours are plotted from tan (lowest population) to
      red (highest population). Different CCD integration times were
      used for recoding these data sets: (A) 25 msec, (B) 50 msec, and
      (C) 100 msec. For interpretation of the significance of these TDPs,
      {\it cf.} Sec.~\ref{sec:results}.
    }
    \label{fig:TDPs}
  \end{center}
\end{figure}

\end{document}